\documentclass[a4paper,noshowpacs,noshowkeys,floatfix,twocolumn,preprintnumbers,nofootinbib]{revtex4-2}
\usepackage{color,amsmath,amsfonts,graphics,epsfig,amssymb}
\usepackage{multirow}
\usepackage{url}
\usepackage{hyperref}
\makeatletter
\renewcommand{\p@subsection}{}
\renewcommand{\p@subsubsection}{}
\makeatother
\begin{document}
\title{
{~ \hfill \rm INR-TH-2021-025}\\
~\\
Constraints on the models of the origin of high-energy astrophysical
neutrinos}
\thanks{Invited review published in \textit{Physics Uspekhi} in a special
issue dedicated to the 50th anniversary of INR RAS, Russian
doi:10.3367/UFNr.2021.09.039062, English
doi:10.3367/UFNe.2021.09.039062.}
\date{\protect\small Received July 2, revised September 7, accepted
September 13, 2021 }
\author{Sergey Troitsky} \email{st@ms2.inr.ac.ru}
\affiliation{Institute for Nuclear Research of the Russian Academy of
Sciences, 60th October Anniversary Prospect 7a, Moscow 117312, Russia}
\begin{abstract}
The existence of astrophysical neutrinos with energies of tens of TeV and
higher has been reliably established by the IceCube experiment; the first
confirmations of this discovery are being obtained with the ANTARES and
Baikal-GVD facilities. The observational results do not fully agree with
what was expected before the start of these experiments.
The origin of these neutrinos has not been
conclusively established, and simple theoretical models,
popular for decades, cannot explain all observational
data. This review summarizes the experimental results with emphasis on
those important for
constraining
theoretical models, discusses  various scenarios
for the origin of high-energy neutrinos and briefly lists
particualr classes of their potential astrophysical sources. It is
demonstrated that the observational data may be explained if the flux of
astrophysical neutrinos includes the contribution of extragalactic sources,
dominating at the highest energies, and the Galactic component, significant
only at neutrino energies $\lesssim 100$~TeV. Other possible scenarios are
also discussed.
\end{abstract}
\maketitle

\thispagestyle{empty}
\newpage
\setcounter{page}{1}
\tableofcontents


\section{Introduction}
\label{sec:intro}
\subsection{Astrophysical high-energy neutrinos: formulation of
the problem}
\label{sec:intro:intro}
Modern astrophysics has confidently moved beyond the so-called photon
channel, the study of sources based on the electromagnetic radiation of
various bands coming from them. The first extraterrestrial sources of
neutrinos, the Sun and supernova 1987A, and then also gravitational waves
have been detected. Extraterrestrial charged particles, cosmic rays, are
being studied intensively. Along with the development of
electromagnetic astronomy, including the highest energies at which
telescopes detect individual photons, this gave rise to the so-called
multimessenger astrophysics, which uses different carriers -- photons,
neutrinos, charged particles and gravitational waves -- to  obtain
information about the structure of astrophysical objects and about
physical processes going on there. Here we will focus on one of the
channels of the multimessenger astronomy, detection of high-energy
neutrinos, and on related observations in other channels.

Due to the unique place of the neutrino among elementary particles (a
stable particle experiencing weak interactions only), the main
goal of the neutrino astronomy in the 20th century was the study of
sources opaque to electromagnetic radiation. Thus, the discovery of solar
neutrinos experimentally proved that the energy of the Sun comes from
thermonuclear reactions in its central region not accessible to other
observations, and the registration of neutrinos from the supernova 1987A
allowed one to verify understanding of physical processes taking place in
the interior of a massive star at the gravitational collapse of its core.
In both cases neutrinos are born in nuclear processes and have energies
corresponding to the characteristic nuclear scales (from fractions to tens
of MeV). Occurring in substantially opaque regions, these processes
``heat'' the source and are eventually related to the thermal
electromagnetic radiation -- for example, the solar radiation. At the same
time a large fraction of the photon radiation in the Universe is
associated with non-thermal processes and is determined by the interaction
of relativistic particles with ambient fields, matter and radiation. As a
rule, such nonthermal emission at relatively low energies, from radio
to ultraviolet, and sometimes up to the X-ray band, is well explained by
the synchrotron radiation of relativistic electrons. At higher energies,
the situation becomes less clear -- along with the synchrotron radiation of
electrons, the inverse-Compton radiation may be significant, as well as the
proton synchrotron or photon production in elementary-particle
interactions. Since the only way to produce astrophysical neutrinos with
energies $\gtrsim 10$~GeV that does not involve non-standard physics or
astrophysics is the interaction of high-energy protons (see
section~\ref{sec:general:pi-mesons}), the role of neutrinos in astronomy
changes with the transition to high energies  -- instead of being carriers
of information about processes in opaque media, they become markers of
interactions of relativistic hadrons whose acceleration to high energies
requires the medium to be not too dense.

This transition from neutrinos born in nuclear processes to neutrinos
associated with high-energy elementary particle interactions is crucial
in distinguishing high-energy neutrino astrophysics as a separate field,
to which this review is devoted. Note that one article, even a large one,
cannot fully cover all aspects of this actively developing area. This
review, therefore, does not pretend to be complete. We will concentrate on
astrophysical models of the origin of neutrinos with energies in the
$(10^{11}\dots 10^{16})$~eV range and will therefore focus only on the
most relevant experimental results for their study. Beyond the scope of
this work are, in particular, ultrahigh energy neutrinos, interesting
details of the experimental work on neutrino detection and various
results in a certain way related to the study of elementary-particle
properties. Even within this framework, a review
of the literature will be necessarily incomplete, for which the author
apologizes in advance. Books and reviews \cite{Capone, Spurio-book,
VissaniUniverse, SpieringUFN, Spiering:reference2020, RyabovUFN} and
others, that touch on a variety of aspects of high-energy neutrino
astronomy, can be recommended to the reader.

\subsection{High-energy neutrino detection}
\label{sec:intro:detection}
\paragraph{Neutrino interactions in water.} The
experimental data to be discussed in this article were obtained using
neutrino telescopes that record Cherenkov emission of charged particles,
the products of neutrino interactions in large volumes of water (in
the solid or liquid state). Interactions of neutrinos with quarks $q$ of
nucleons of target atomic nuclei can proceed with the $W$-boson exchange
(charged current, CC),
\begin{equation}
\mbox{CC:}~\nu_{l}+q\to l+X,
\label{Eq:CC}
\end{equation}
where $\nu_{l}$ and $l$ are neutrinos and charged leptons of the same
flavor, $l=e,\mu,\tau$, and $X$ denotes other, hadronic
products of the reaction. The other interaction channel is the $Z$-boson
exchange (neutral current, NC),
\begin{equation}
\mbox{NC:}~\nu_{l}+q\to \nu_{l}+X.
\label{Eq:NC}
\end{equation}
Similar reactions are possible for antineutrinos.

The probability of interaction between neutrinos and target electrons is
low except in the case of the so-called Glashow
resonance~\cite{Glashow:1960zz, BerezinskyGazizov-GlashowRes}, the direct
production of a $W$-boson,
\begin{equation}
\bar\nu_{e}+e\to W\to\dots,
\label{Eq:Glashow-res}
\end{equation}
where ``$\dots$'' denotes a well-studied set of $W$-boson decay products,
which can include both hadrons and leptons. This process goes only for
antineutrinos, for there are no positrons in the target. The resonance
occurs at the energy $E_{\bar\nu_{e}}=m_{W}^2/(2m_{e})$ where $m_{W}$ and
$m_{e}$ are the masses of the $W$ boson and electron, respectively.

The result of the interactions observed in the detector depends not only
on the type (\ref{Eq:CC}), (\ref{Eq:NC}), (\ref{Eq:Glashow-res}), but
also on the flavor of the initial neutrino. The CC reaction involving
$\nu_{\mu}$ ($\bar\nu_\mu$) results in the production of a relativistic
muon, which decay length is, at the energies of interest,
usually larger than the size of the detector. The Cherenkov radiation of
this single muon is recorded as a narrow \textbf{track} crossing the
detector. Note that the widespread notion that such track events are
associated only with $\nu_{\mu}$ is not entirely correct: they also
include muons from the decay of $\tau$-leptons born in CC interactions of
$\nu_{\tau}$, or from $W\to \mu^{-} \bar\nu_{\mu}$ in the case of
the Glashow resonance, as well as high-energy $\tau$-leptons that do not
have time to decay in the detector (a total of $\sim 10\%$ of all
tracks~\cite{IC7yrPoint-tracksNotOnlyMu}).

CC events involving $\nu_{e}$ or $\bar\nu_{e}$ lead to the formation in
the detector of two multiparticle processes, overlapping
and forming a common \textbf{cascade}. One of the showers starts with the
braking radiation of the electron, whose radiation length in water is only
about 36~cm, and the other is associated with the hadronic products of
$X$. Note that the longitudinal development of the cascade occurs over a
length three orders of magnitude shorter than the length of a muon track
(see, e.g., estimates in Ref.~\cite{VissaniUniverse}).

The CC reaction caused by $\nu_{\tau}$ or $\bar\nu_{\tau}$ looks
differently in the detector depending on the neutrino energy. The decay
length of a $\tau$-lepton with energy $E_{\tau}$ is $\sim
50~\mbox{m}\times \left(E_{\tau}/\mbox{PeV} \right)$, so at PeV energies
such event is observed as a \textbf{double
cascade}~\cite{Learned-DoubleBang} -- at the interaction point, a shower
of $X$ hadrons is recorded, and at the decay point -- one from $\tau$ decay
products, also predominantly hadronic. It is possible to separate the
showers both spatially and by the time interval between the two
flashes \cite{IceCube3yrDoublePulse}. At the energies of $\lesssim$PeV
these interactions look like  normal cascades.

Finally, at NC interactions of neutrinos and antineutrinos of all types
only the hadronic cascade is recorded, as the neutrino remaining in the
final state leaves the detector without further interactions.

\paragraph{Specifics of observation of track and cascade events.}
Events interpreted as the Glashow resonance or a double cascade are
associated with the highest neutrino energies and thus are very
rare~\cite{IceCube2021GlashowRes, IceCube2bangObs}, so the bulk of
astrophysically interesting information is associated with tracks and
cascades. In practice, the registration of Cherenkov radiation is made by
a three-dimensional array of optical modules with photodetectors viewing a
large volume of the target. In order to reduce the background from muons
of extensive atmospheric showers (see later in this section), the facility
is immersed in water (ice) to a depth of about one kilometer.
Photodetectors record the amount of Cherenkov light and, with high
accuracy, the moment of the flash. The latter is important because the
temporal evolution of the signal makes it possible to determine the
direction and speed of a muon or of the cascade development. Illustrations
with images of simulated and real recorded events can be found in the
reviews cited above, popular science literature (see, e.g.,
\cite{ST-mnogok}) and the media.

The cascade evolves almost isotropically in the center-of-mass system of
the original particles, so in the laboratory frame it looks
like an extended ``cloud''. The neutrino arrival direction is
determined considerably worse in this case than for a long
linear muon track, see examples below. In contrast, in terms of the energy
determination, cascades are indispensable: if a cascade begins in the
working volume of the detector (the so-called starting events), then
almost all of the energy of the initial neutrino goes to the Cherenkov
light and is collected by photodetectors.

\begin{figure}
\centerline{\includegraphics[width=\columnwidth]{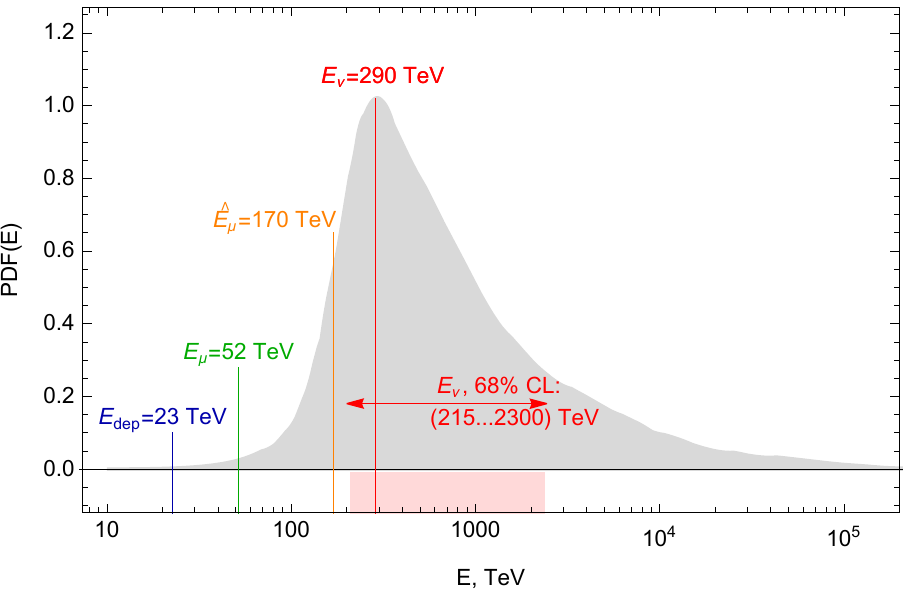}}
\caption{
\label{fig:E-uncert}
An illustration of the energy-estimate uncertainty
for a track event using the example of one of the best known
neutrinos registered by IceCube (IC170922A, coincident with
the outburst of the blazar TXS~0506$+$056, see
Sec.~\ref{sec:exper:aniso:flares}). Horizontal axis -- energies: the energy
deposited in the detector $E_{\rm dep}=23.7\pm2.8$~TeV, the
reconstructed energy of the muon entering the detector
$E_{\mu}=52^{+11}_{-9}$~TeV, the energy estimate of the muon at birth $\hat
E_{\mu}\simeq 170$~TeV, the most likely neutrino energy $E_{\nu}\simeq
290$~TeV. The shaded graph shows the probability density function (PDF) of
the values of $E_{\nu}$; also shown is the uncertainty region of the
$E_{\nu}$ values  -- from 215 to 2300~TeV (68\% CL). A power-law spectrum
of astrophysical neutrinos with the index of 2.13 is assumed. Plotted on
the basis of data from Ref.~\cite{IceCube-ScienceTXS1}.}
\end{figure}

For track events, the potential of determining the energy of the initial
neutrino is noticeably more modest. The muon track usually extends beyond
the boundaries of the facility, so the total recorded energy release in
the detector, $E_{\rm dep}$, gives only a lower bound on the muon
energy $E_{\mu}$ at the entrance to the detector, which is estimated using
the time evolution of the signal. Since it is not known how much energy
the muon has lost before entering  the detector, as well as how
much of the energy of the original neutrino was transferred to the muon,
this measurement allows one to estimate the energy of the muon at birth
$\hat E_{\mu}$ and the energy of the original neutrino $E_{\nu}$ only
in a statistical way. While $E_{\nu}$ is constrained from below quite well,
the broad non-Gaussian distribution of possible $E_{\nu}$
corresponding to a given muon detection,
extends towards
values of $E_{{\nu}}$ orders of magnitude higher
(see
Fig.~\ref{fig:E-uncert}).
Additionally, note that this statistical distribution, as well as
the most likely estimate of $E_{\nu}$, depends on the assumption about the
spectrum of astrophysical neutrinos.

\paragraph{Arrival directions. Water and ice.}
In particle-physics terms, the interactions of neutrinos in liquid
water and in ice are of course identical, but for the
detection of Cherenkov radiation and for event reconstruction,
properties of these media are quite different.

Table~\ref{tab:water-ice}
\begin{table*}
\begin{center}
\begin{tabular}{cccc}
\hline
\hline
 & Lake Baikal & Salty & Ice\\
 & water       & water & \\
\hline
Absorption length, m & $\sim 20$ & $\sim 50$ & $\sim 100$ \\
Scattering length, m & $\sim 200$ & $\sim 200$ & $20-40$ \\\
Problems &
\begin{minipage}{3.3cm}
Shallower available depth, bioluminescence, chemiluminescence
\end{minipage}
&
\begin{minipage}{3.3cm}
Radioactivity of dissolved salts, bioluminescence, chemiluminescence
\end{minipage}
&
\begin{minipage}{3.3cm}
Dust inclusions, clathrates
\end{minipage}
\vspace{1mm}
\\
\hline
\hline
\end{tabular}
\end{center}
\caption{\label{tab:water-ice}
A comparison of water and ice detectors.
Longer length corresponds to better optical performance.}
\end{table*}
gives characteristic values of the absorption and scattering lengths
of light at
wavelengths of the maximal Cherenkov radiation for typical
experimental conditions in different media (for details
see Ref.~\cite{ChiarusiSpurio2009}). Under real conditions, these values
depend strongly on the particular location, primarily on the depth, varying
even within the same installation. Nevertheless, the averaged
estimates show that the instruments using liquid water and ice
complement each other. In terms of identifying events and measuring their
energies, ice is more convenient -- it has less illumination from natural
sources (bio- and chemiluminescence and radioactivity), and the weak
absorption allows one to collect more light from each
event, lowering the registration threshold and increasing the accuracy of
the energy determination. In contrast, for the
astrophysical task of identifying neutrino sources,
the accuracy of the arrival direction, including both the
statistical scatter and systematic errors, is of more importance. The
statistical errors are largely determined by
the scattering length, so that they are about 5 times smaller in water
experiments than in ice experiments; this is true for both cascades and
tracks. In the case of cascade events this improvement (from
$15^{\circ}-20^{\circ}$ to $3^{\circ}-4^{\circ}$) is crucial.

\begin{figure}
\centerline{\includegraphics[width=\columnwidth]{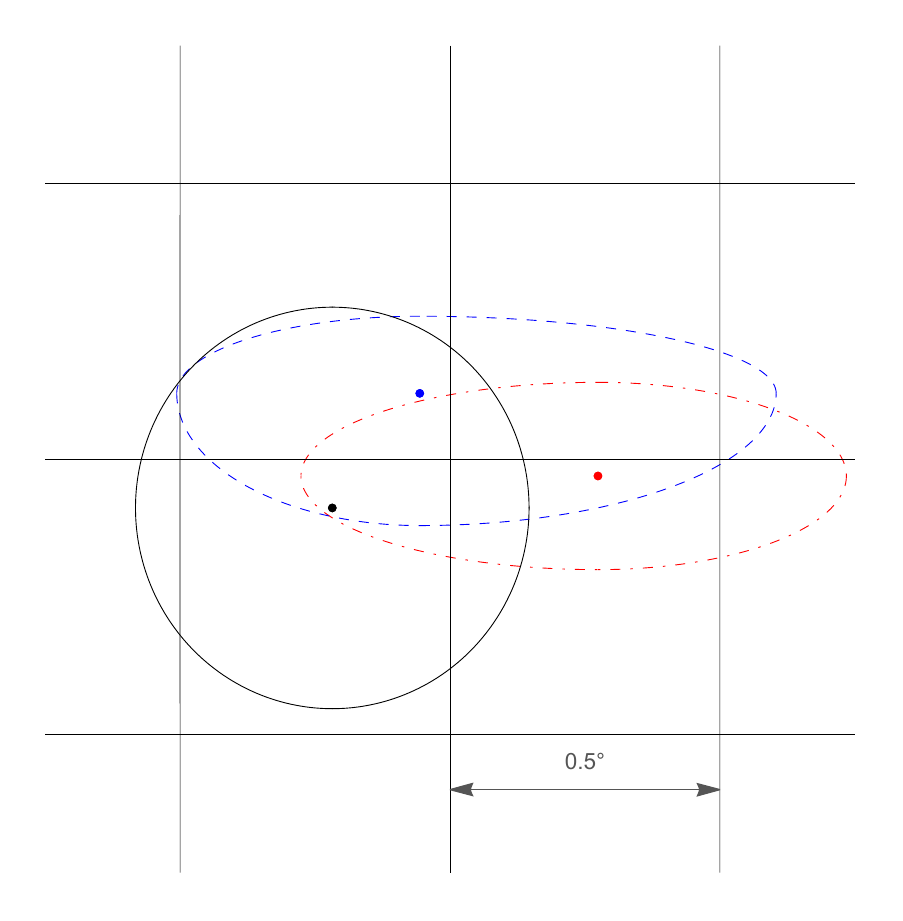}}
\caption{
\label{fig:recon-uncert1}
Illustration of statistical and systematic
uncertainties of arrival directions of track events. The arrival
directions and their 90\% CL statistical error regions are shown for the
same event with energy $\sim$4450~TeV, according to
Ref.~\cite{IceCube-mu2016} (dashed dashed line), the catalog of alert
events \cite{IceCubeEHEAcatalog} (dashed line) and the IceCube 10-year
public catalog \cite{IceCube10yrData, IceCube10yrDataPaper} (solid
line). The difference between reconstructions gives an estimate of
the systematic uncertainty.}
\end{figure}

\begin{figure}
\centerline{\includegraphics[width=\columnwidth]{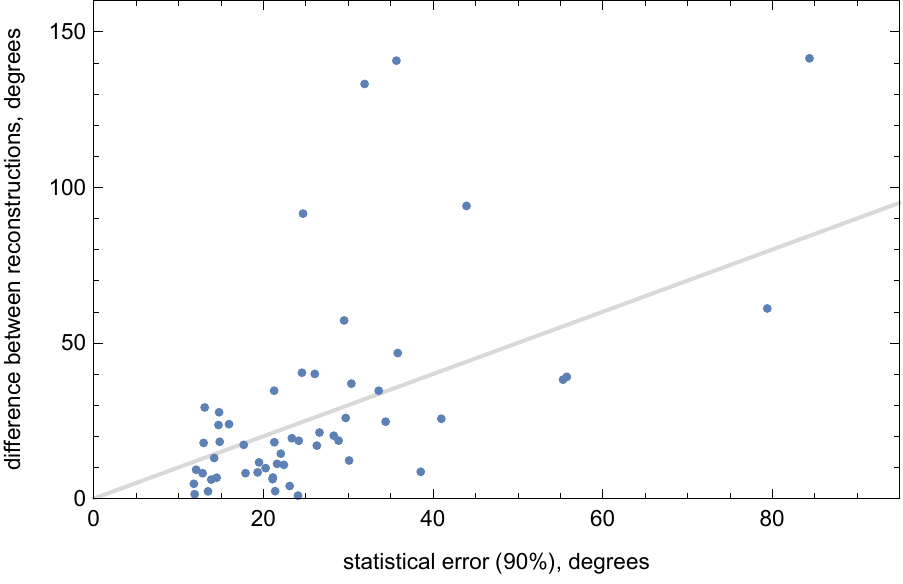}}
\caption{
\label{fig:recon-uncert2}
Illustration of statistical and systematic
uncertainties of arrival directions of cascade events.
The horizontal axis gives the value  of
the statistical error (90\% CL) of the arrival direction of HESE cascades
of Refs.~\cite{IceCube-HESE1, IceCube-HESE2, ps1710.01179p54}; the vertical
axis gives the difference between the arrival directions of these events
in the original \cite{IceCube-HESE1, IceCube-HESE2, ps1710.01179p54} and
new \cite{HESE2020} reconstructions.}
\end{figure}

Another important component of the accuracy of the reconstruction of
neutrino arrival directions are systematic uncertainties. These
come primarily from  the accuracy of positioning the
detector as a whole and from imperfections in the reconstruction of
events, including insufficient knowledge of the properties of the medium.
Since inhomogeneities, bubbles and inclusions in water mix, float or
settle down, the working volume of a water detector is much more
homogeneous and much more uniform and controllable than that of an ice
detector. If the accuracy of absolute positioning can be estimated from
observation of the shadow of the Moon, the second component of
the systematic uncertainties is much harder to estimate. Some estimates of
its magnitude may be obtained by comparing the arrival directions of the
same events reconstructed with different algorithms and with
different ice models, see Figs.~\ref{fig:recon-uncert1},
\ref{fig:recon-uncert2},
where also some typical statistical errors in determination of the neutrino
arrival directions are shown.

\begin{figure}
\centerline{\includegraphics[width=\columnwidth]{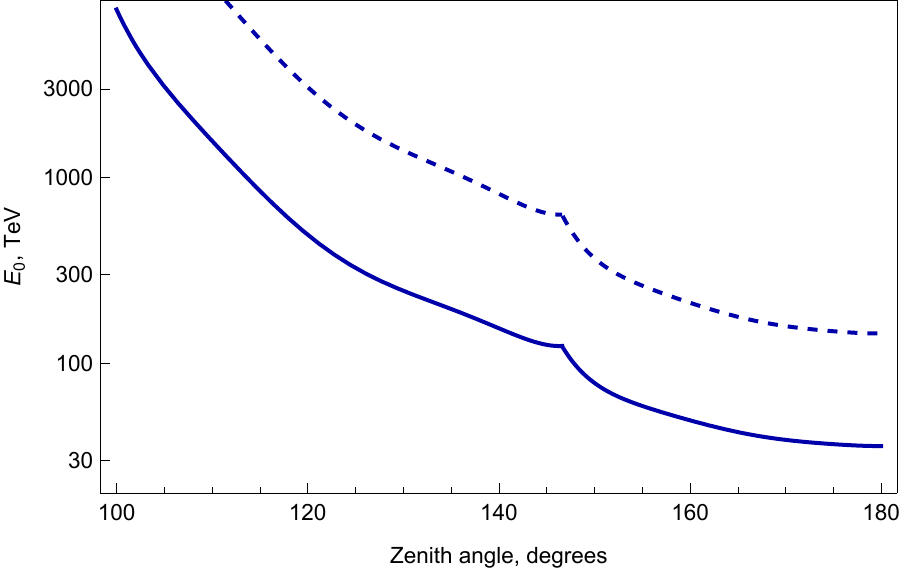}}
\caption{\label{fig:earth}
The critical energy $E_{0}$ for which the optical depth for
the electron neutrino with respect to its interaction with the Earth's
matter is 1 (solid line) or 2.3 (dashed line,  90\% of neutrinos
interact) as a function of the zenith angle. At energies $\gtrsim E_{0}$,
the Earth gradually becomes opaque to neutrinos coming from this
direction. }
\end{figure}

\paragraph{Passage of neutrinos through the Earth.}
Neutrino interaction cross sections increase with energy, and at the high
energies of interest, the Earth is no longer completely transparent to
neutrinos. For different arrival directions,
paths through the Earth are different, but the dependence of the
interaction probability is not just geometric because the Earth has a very
dense core with a sharp boundary. This issue is discussed in more detail
in Refs.~\cite{Capone, VissaniUniverse}; here we present only
Fig.~\ref{fig:earth}, plotted on the basis of the data from these works.
It shows the characteristic neutrino energy, from which the interaction
with the Earth is significant, as a function of the zenith angle. Note
that some of the interactions are elastic, after which the neutrino
continues its motion with lower energy, so that the total flux suppression
depends also on the spectrum of the incoming particles, not only on the
zenith angle.

\begin{figure}
\centerline{\includegraphics[width=\columnwidth]{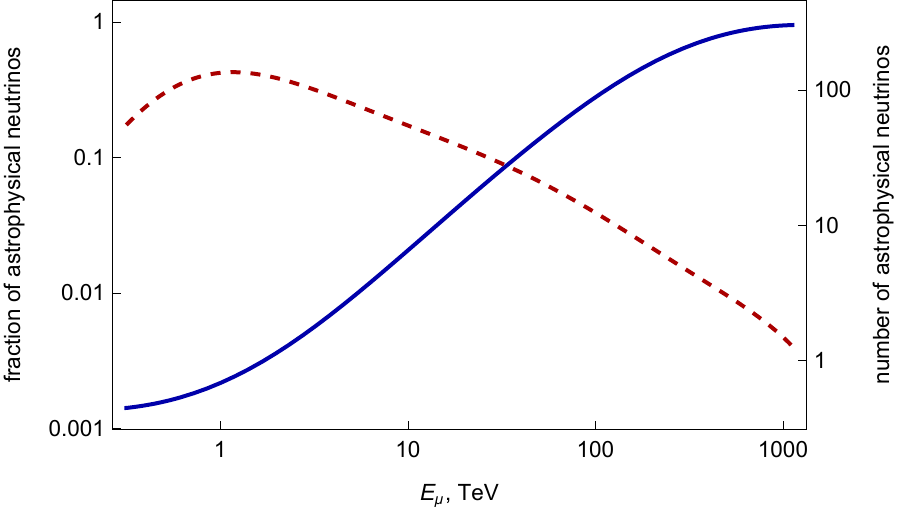}}
\caption{\label{fig:astro-fraction}
The expected fraction of astrophysical neutrinos in the total number of
muon tracks  as a function of energy $E_{\mu}$ (left scale, solid line)
and the expected total number of astrophysical neutrinos at a given energy
(per 0.1 dex bin in $E_{\mu}$, right scale, dashed line) in a set of $\sim
650000$ IceCube events. Plotted with the data from
Ref.~\cite{IceCube-mu2019} (model for 10 years of observations). }
\end{figure}

\paragraph{Atmospheric and astrophysical neutrinos.}
The main backgrounds for detection of astrophysical neutrinos are muons and
neutrinos from interactions of cosmic rays with the Earth's atmosphere.
Except that at the highest energies, this background dominates over the
signal. For instance, Fig.~\ref{fig:astro-fraction}
gives an estimate of the fraction of astrophysical events in the set of
IceCube muon tracks as a function of the muon energy $E_{\mu}$. Despite
such a modest signal-to-background ratio, there are ways to isolate the
astrophysical component in the analysis.

For energies of the order of GeV, the process of atmospheric neutrino
production in interactions of cosmic rays with hadrons present in the
atmosphere is quite similar to the mechanism of the origin of astrophysical
neutrinos in $pp$ interactions, see Sec.~\ref{sec:general:pi-mesons} below:
$\pi^{\pm}$ mesons are born and then decay, and then the $\mu^{\pm}$,
produced at this first stage, decay as well. The resulting neutrino
spectrum reflects the spectrum of cosmic rays, $E^{-2.7}$, and the flavor
ratio $\nu_{e}:\nu_{\mu}:\nu_{\tau}$ is 1:2:0. The situation changes for
the muon energies $\hat E_{\mu} \gtrsim 10$~GeV, when muons begin to
reach to the Earth's surface without decay. As a result, the flavor
composition changes so that the ratio of $\nu_{e}:\nu_{\mu}$ becomes $\sim
1:30$ (electron neutrinos are born in some $K$-meson decays; $\nu_{\tau}$
still has no mechanism to be produced). In addition, at the $\pi$-meson
energies  $E_{\pi} \gtrsim 100$~GeV, the latter also have no time to decay
-- they interact faster  with atmospheric hadrons;
atmospheric showers begin to develop. Since the decay probability falls
with energy as $1/E_{\pi}$ (Lorentz kinematics), and the interaction
probability weakly depends on the energy, in the energy region of
interest, the spectrum of atmospheric neutrinos from  $\pi$- and
$K$-meson decays follows $\sim E^{-3.7}$ (see
Sec.~\ref{sec:exper:spec-flav} below for the discussion of the
contribution of charmed hadrons).

The distribution of atmospheric neutrinos in zenith angles peaks strongly
for horizontal directions, since in this case the muon's path in the
atmosphere is longer, and hence it is more likely to decay. In real
analyses at high energies, this dependence is further strengthened by
applying muon veto for events coming from above (events with simultaneous
signal from other muons from the same shower are rejected), while for
$E_{\nu} \gtrsim 50$~TeV the Earth's opacity to neutrinos starts to be
noticeable for events coming from below. It is the combination of the very
soft spectrum $E^{-3.7}$ and the described dependence on the zenith angle
which gives  the basis for the extraction of the astrophysical
signal against the atmospheric background on the statistcal basis, see
Figs.~\ref{fig:atm-astro-E}, \ref{fig:atm-astro-ZA} in
Sec.~\ref{sec:exper:extraterr} below.

\subsection{Past, present, and future experiments}
\label{sec:intro:hist}
The idea of underwater detection of high-energy neutrinos was first
proposed by M.A.~Markov and I.M.~Zheleznykh~\cite{Markov:1960vja}
(see also Ref.~\cite{Zheleznykh}).
It is not the purpose of this review to
discuss in detail the history of neutrino astronomy, nor to give a detailed
technical description of the instruments at work, see
Refs.~\cite{Spurio-book, SpieringUFN, Spiering:reference2020} and
references therein. Brief information about past, present, and emerging
detectors, which may be useful in reading the rest of the review, can be
found in Table~\ref{tab:experiments}.
\begin{table*}
\begin{center}
\begin{tabular}{ccccl}
\hline
\hline
Name & Location & Volume, km$^{3}$ & Years & Note \\
\hline
NT-36,&&&&First detection \\
 NT-200,&Baikal&$10^{-4}$ (*)&1993--2015&of muon tracks from\\
 NT-200+&&&&atmospheric neutrinos\\
\hline
AMANDA& South Pole&0.015&1996--2008& Atmospheric neutrino\\
&&&& spectrum up to $\sim 100$~TeV\\
\cline{1-4}
ANTARES&Mediterranean&0.025 (*)&2006--...&and constraints on \\
&sea&&&astrophysical models\\
\hline
\hline
&&&& Observation of astrophysical\\
IceCube&South Pole&1.0&2006--...& neutrinos. The largest\\
&&&& statistics (2021)\\
\hline
Baikal-GVD&Baikal&0.4 (2021) (*)&2016--...&Data taking \\
&&$\ge 1$ (plan)&& in progress \\
\cline{1-4}
KM3NeT&Mediterranean sea&$\sim 1$ (*), plan&2019--...&of deployment\\
\hline
\hline
IceCube-Gen2&South Pole&$\sim 10$, plan&project&\\
\hline
P-ONE&Pacific, Canada&$\sim 3$(*), plan&project&\\
\hline
\hline
\end{tabular}
\end{center}
\caption{\label{tab:experiments}
Major experiments in high-energy neutrino astrophysics. Specified are
the years of obtaining physical results, including those in incomplete
configurations.  (*) the volume filled with detecting
equipment; water detectors can be used to record
high-energy cascades in the volume well in excess of these values.}
\end{table*}

\section{Key experimental results}
\label{sec:exper}
\subsection{Extraterrestrial origin of neutrinos}
\label{sec:exper:extraterr}
\begin{figure}
\centerline{\includegraphics[width=\columnwidth]{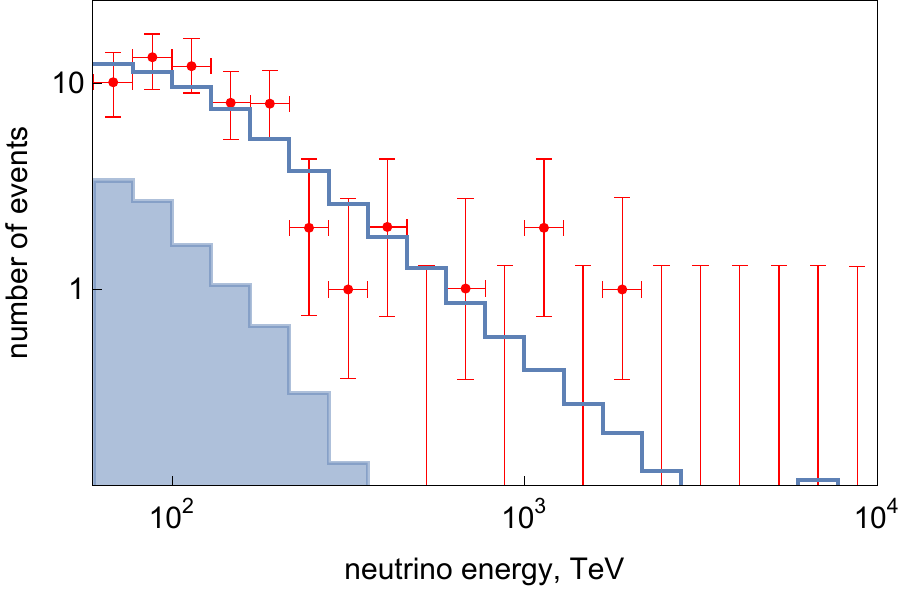}}
\caption{\label{fig:atm-astro-E}
Distribution of IceCube cascade events, starting in the detector, in the
energy deposited in the detector ($E>60$~TeV). Red dots with error bars --
data, dark shading -- standard atmospheric background, solid line
-- the fit of the sum of the background and the astrophysical component.
Plotted with the data from
Ref.~\cite{HESE2020}.
}
\end{figure}

\begin{figure}
\centerline{\includegraphics[width=\columnwidth]{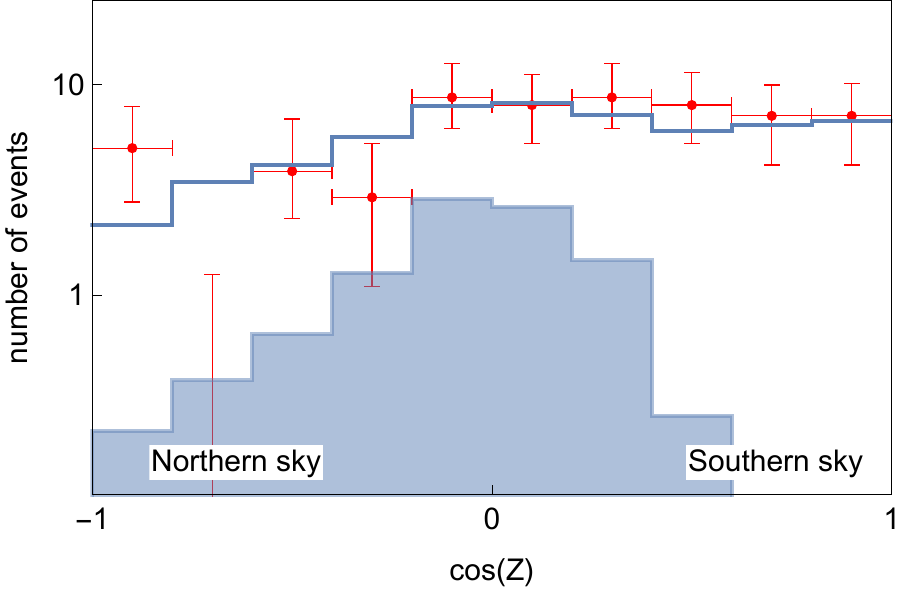}}
\caption{\label{fig:atm-astro-ZA}
Distribution of IceCube cascade events ($E>60$~TeV), starting in the
detector, in the zenith angle $Z$. Red dots with error bars
-- data, dark shading -- standard atmospheric background, solid line
-- the fit of the sum of the background and the astrophysical component.
Plotted with the data from Ref.~\cite{HESE2020}.
}
\end{figure}

As it has been already noted, we are discussing neutrinos with energies
substantially above the nuclear scales. There are no terrestrial sources
of such neutrinos (except for narrow beams of neutrinos of accelerator
origin, which however do not enter into experimental facilities of the
cubic-kilometer scale). The astrophysical signal in the detectors should
be separated from atmospheric neutrinos produced by interactions of
cosmic rays with the Earth's  atmosphere, and from background events
caused by muons from the same interactions. When studying a single
neutrino event, it is not possible to say definitely whether
it was of the atmospheric or astrophysical origin, but in examining the
ensemble of the data it is possible to distinguish the astrophysical
component of the flux against the atmospheric one. For this purpose,
one uses the distribution of events in energies (at high energies, hard
astrophysical spectra become more pronounced compared to soft atmospheric
ones, see Fig.~\ref{fig:atm-astro-E})
and the zenith angle (see Fig.~\ref{fig:atm-astro-ZA}).
While astrophysical neutrinos come fairly isotropically
(only at the highest energies does the Earth become opaque to them),
the distribution of atmospheric neutrinos by zenith angles peaks for
horizontal directions.
The atmospheric
muons themselves can be filtered out by a simultaneous triggering system
mounted on the surface, or -- for directions from below -- simply
by the Earth.

Detection of neutrinos of astrophysical origin was first announced
by the IceCube Collaboration in 2013 based on the observation of
two events with cascades started in the detector with reconstructed
energies above 1~PeV, for which the atmospheric background is simply
negligible. Presently, various analyses based on IceCube
datasets with different selection criteria are carried out, and in all of
them the presence of an astrophysical component of the neutrino flux is
established in a statistically significant way, based on a combination of
multiple factors. For the most energetic individual events, the
probability of their astrophysical origin is determined (we
note that for particular samples used in the IceCube analyses, even for
those selected according to the most stringent criteria, this probability
in average does not exceed $\sim 60\%$).

An independent confirmation of the presence of astrophysical
neutrinos by other experiments should be considered as an important step in
the development of neutrino astrophysics. Such a result, albeit with low
statistical significance, was presented in 2019 by the ANTARES
collaboration \cite{ANTARESastrophys_flux}. The number of events with
energies above 100~TeV recorded at the Baikal-GVD facility during data
taking in the incomplete configuration is also
consistent~\cite{Baikal-astro-events} with the presence of the
astrophysical neutrino flux with the parameters measured by IceCube. An
eagerly awaited more accurate quantitative verification of the IceCube
results will be possible in a few years of Baikal-GVD operation.

\subsection{Spectrum and flavor composition}
\label{sec:exper:spec-flav}
Detailed measurements of the spectrum of astrophysical neutrinos, similar
to those carried out for cosmic rays, are still difficult because of
statistical and systematic uncertainties. The energies of individual track
events are estimated with low accuracy which does not allow for any
meaningful binning of the spectrum, while the number of reliably studied
starting cascade events is small. In addition, the estimation of the
astrophysical flux is always based on subtracting the atmospheric
background, which is also modeled with uncertainties. As a consequence,
most of the spectrum estimates obtained so far use a simple power-law fit,
\begin{equation}
\frac{dF_{\nu+\bar\nu}}{dE_{\nu}} = \Phi_{0}
\left(\frac{E_{\nu}}{\mbox{100~TeV}} \right)^{-\gamma} \times
10^{-18}~\mbox{GeV}^{-1}\,\mbox{cm}^{-2}\,\mbox{s}^{-1}\, \mbox{sr}^{-1},
\label{Eq:plaw}
\end{equation}
where the fit parameters are the normalization $\Phi_{0}$ and the spectral
index $\gamma$. The standard para\-metri\-zation (\ref{Eq:plaw}) refers to
the diffuse isotropic flux of neutrinos and antineutrinos of the same
flavor, defined assuming equal fluxes of all six types of neutrinos and
antineutrinos (thus, the total flux is obtained by multiplying that given
by Eq.~(\ref{Eq:plaw}) by three). Different analyses are more or less
sensitive to different energy intervals and different flavors. Parameters
of the best-fit spectrum (\ref{Eq:plaw}), determined in the various
IceCube and ANTARES analyses, are given in Table~\ref{tab:plaw-fits}.
\begin{table*}
\begin{center}
\begin{tabular}{cccc}
\hline \hline Analysis & Energy & $\Phi_{0}$ & $\gamma$ \\
\hline HESE 2020 \cite{HESE2020}& 69.4~TeV--1.9~PeV &
$2.12^{+0.49}_{-0.54}$ & $2.87^{+0.20}_{-0.19}$ \\
Cascades $\nu_{e}+\nu_{\tau}$ 2020 \cite{Cascades2020e-tau}&
16~TeV--2.6~PeV& $1.66^{+0.25}_{-0.27}$ & $2.53\pm0.07$ \\
MESE 2014 \cite{MESE2014}& 25~TeV--1.4~PeV& $2.06^{+0.4}_{-0.3}$ &
$2.46\pm0.12$ \\
Inelasticity 2018 \cite{inelasticity2018}&
3.5~TeV--2.6~PeV& $2.04^{+0.23}_{-0.21}$ & $2.62\pm0.07$\\
$\nu_\mu$ 2016
\cite{IceCube-mu2016}& 194~TeV--7.8~PeV& $0.90^{+0.30}_{-0.27}$ &
$2.13 \pm 0.13$\\
$\nu_\mu$ 2019 \cite{IceCube-mu2019}& 40~TeV--3.5~PeV&
$1.44^{+0.25}_{-0.24}$ & $2.28^{+0.08}_{-0.09}$ \\
ANTARES 2019 \cite{ANTARESastrophys_flux} &&$1.5 \pm 1.0$ & $2.3\pm 0.4$\\
\hline
\hline
\end{tabular}
\end{center}
\caption{
\label{tab:plaw-fits}
Parameters of the power-law fit (\ref{Eq:plaw}) spectra
of astrophysical neutrinos from various IceCube and preliminary
ANTARES analyses. }
\end{table*}
The ranges of
energies, which give the main (usually 90\%)
contribution to the fit spectrum, are also indicated. A more detailed
discussion of the features of the datasets used and of details of the
experimental work is beyond the scope of this paper; a rather
detailed description of different IceCube analyses is given in a
recent paper~\cite{HESE2020}.

\begin{figure}
\centerline{\includegraphics[width=\columnwidth]{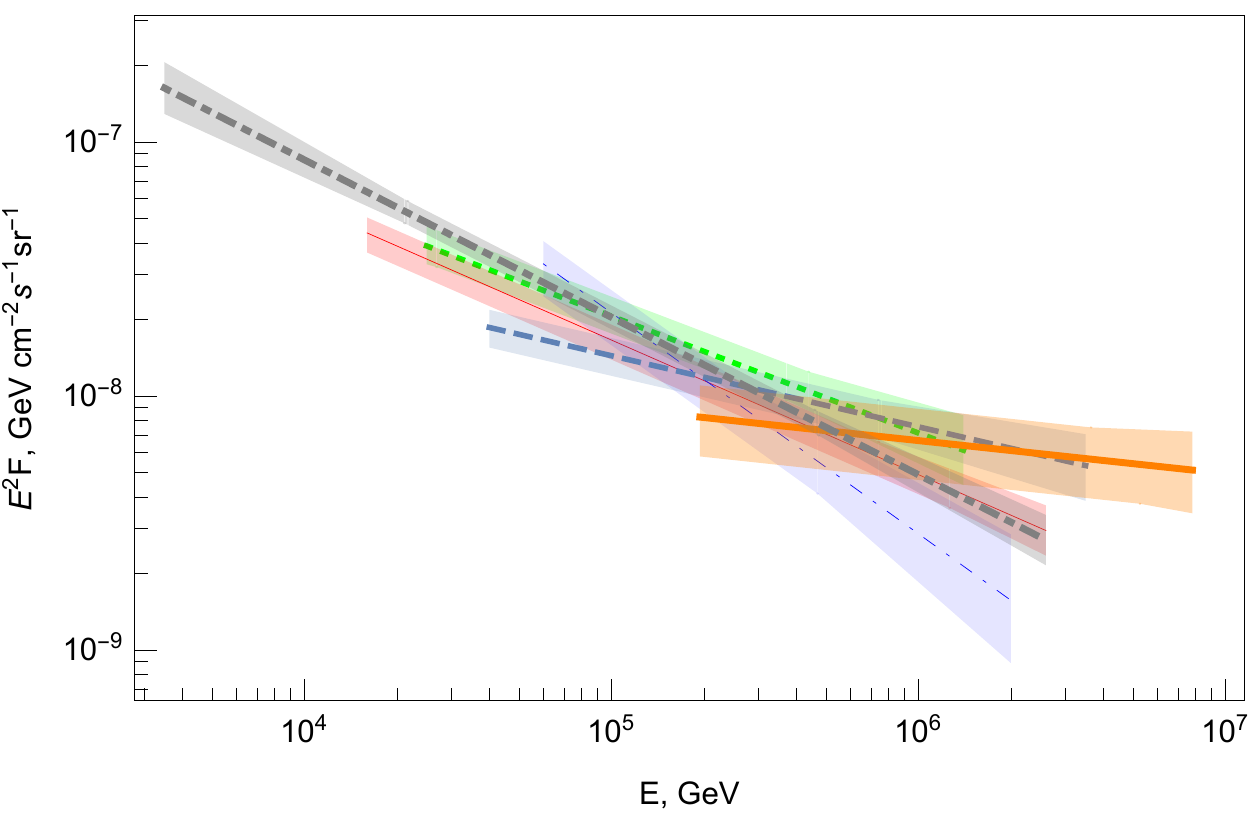}}
\caption{\label{fig:tyes}
Astrophysical neutrino spectra (\ref{Eq:plaw}) from
various IceCube analyses (for details and references, see
Table~\ref{tab:plaw-fits}).
Thick orange solid line -- $\nu_{\mu}$ 2016;
blue dashed line -- $\nu_{\mu}$ 2019;
green dotted line -- MESE 2014;
thick gray dot-dashed line -- inelasticity 2018;
thin red solid line -- $\nu_{e}+\nu_{\tau}$ cascades 2020;
thin blue dot-dashed line -- HESE 2020.
The shading of the corresponding color shows the statistical
uncertainties of the corresponding power-law fit. }
\end{figure}

\begin{figure}
\centerline{\includegraphics[width=\columnwidth]{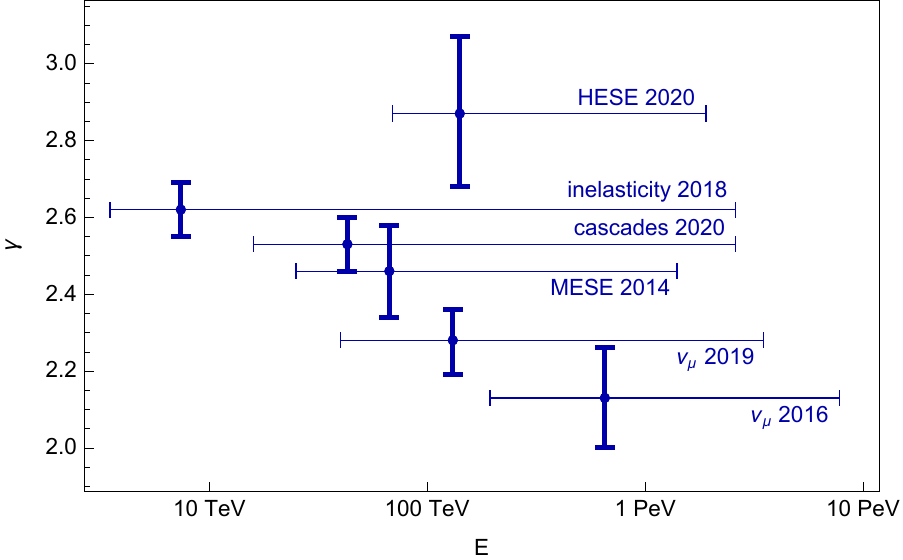}}
\caption{\label{fig:meanE-index}
The index of the power-law fit (\ref{Eq:plaw}) of the astrophysical
neutrino spectrum for various IceCube analyses (for the names of
the analyses and references, see Table \ref{tab:plaw-fits}). Bold
vertical lines show statistical uncertainties of the reconstructed
exponents, and thin horizontal lines show the neutrino energy intervals
yielding 90\% of the events for this analysis. The horizontal positions
of the dots correspond to the average energy of the events in the data
set, estimated from the power-law fit. }
\end{figure}

Although the parameters of the spectra obtained in the different analyses
are close to each other in the order of magnitude, their scatter
is strikingly greater than the 68\% confidence intervals indicated in the
table. Errors of the parameters $\Phi_{0}$ and $\gamma$ are not completely
independent, so the contours  of the confidence regions in the plane
of these two parameters are often compared, see e.g, Ref.~\cite{HESE2020}.
Such a comparison does not always seem to be optimal due to the fact that
the energy intervals used in the analyses are different, and the true
spectrum is probably different from the exact power law. This latter
possibility is supported by a visual comparison of the power-law fitted
spectra plotted for the energy ranges of the respective analyses, see
Fig.~\ref{fig:tyes}.
It can be seen that, on average, the spectra become harder at higher
energies, and at a fixed energy the agreement between those analyses
which have sufficient statistics in this region are not bad. The neutrino
fluxes fall rapidly with energy in any case, so the best statistics
saturating the fit is obtained closer to the lower boundary of the energy
range in use. It is natural to attribute the ``measurement'' of the
power-law index $\gamma$ to the average energy of events contributing to
the fit, which is easy to estimate by knowing the $\gamma$ value itself
and the energy range. The results of such estimation are shown in
Fig.~\ref{fig:meanE-index}.
The tendency for the spectrum to become harder with increasing energy is
particularly clear in this representation --  only one point
\cite{HESE2020} drops out, which, however, is based on only 60
events, albeit of high quality. Possible reasons for this behavior of the
spectrum deserve a more detailed discussion, to which we now turn.

\begin{figure}
\centerline{\includegraphics[width=\columnwidth]{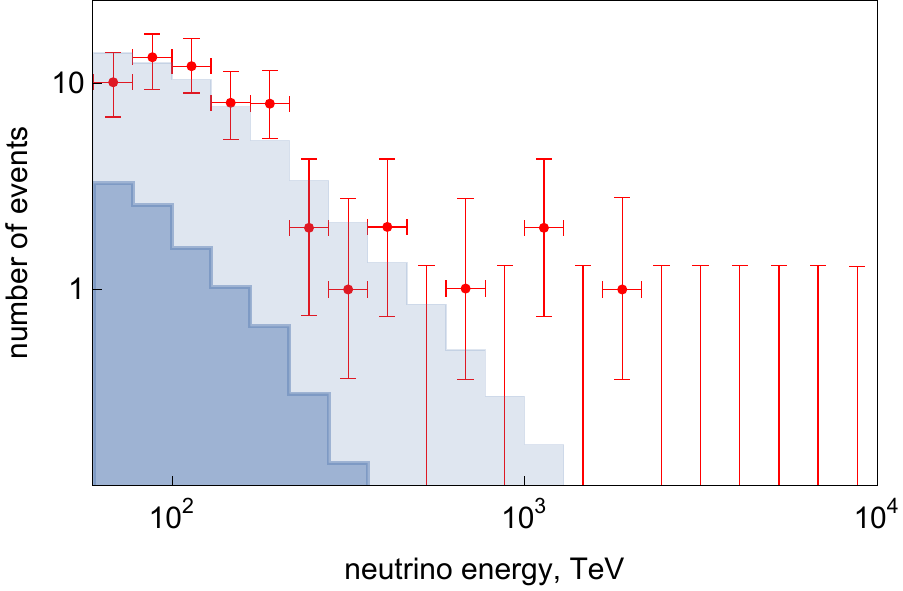}}
\caption{
\label{fig:atm-prompt-E}
Distribution of IceCube cascade events, starting in the detector, in the
energy deposited in the detector ($E>60$~TeV). Red dots with error bars --
data, dark shading -- standard atmospheric background,
light shading -- the best fit of the sum of the standard background
and the background of prompt atmospheric neutrinos.
Plotted with the data from
Ref.~\cite{HESE2020}.}
\end{figure}

\begin{figure}
\centerline{\includegraphics[width=\columnwidth]{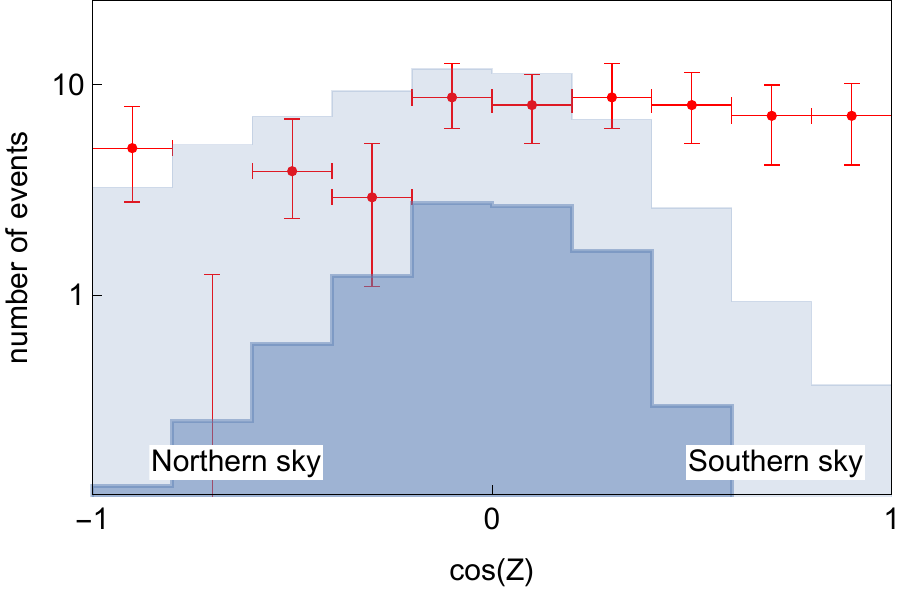}}
\caption{
\label{fig:atm-prompt-ZA}
Distribution of IceCube cascade events, starting in the detector, in the
zenith angle $Z$. Red dots with error bars --
data, dark shading -- standard atmospheric background,
light shading -- the best fit of the sum of the standard background
and the background of prompt atmospheric neutrinos.
Plotted with the data from
Ref.~\cite{HESE2020}.}
\end{figure}

\paragraph{Atmospheric background?}
\label{sec:exper:spec-flav:prompt}
As discussed above, isolating the contribution of astrophysical neutrinos
from the background atmospheric neutrinos at high energies is quite easy:
astrophysical neutrinos have a harder spectrum and a different
distribution of zenith angles, see  Fig.~\ref{fig:atm-astro-E},
\ref{fig:atm-astro-ZA}. However, this refers to so-called standard
atmospheric neutrinos from decays of $\pi$ and $K$ mesons. At the same
time, the products of the interaction of cosmic rays with the
atmosphere include charmed
mesons, $D_{s}^{\pm}$, and baryons $\Lambda_{c}$, whose lifetime is about
four orders of magnitude shorter. Their decays produce the so-called
prompt atmospheric neutrino flux, which has characteristics much
closer to those of the astrophysical flux. In particular, due to the short
lifetime, the parent hadrons have no time to
interact in the atmosphere even at high energies. As as a consequence,
prompt neutrinos repeat the spectrum of the initial cosmic rays,
$E^{-2.7}$, and are distributed isotropically, up to the effect of passing
through the Earth and of event selection. Compared to ordinary
atmospheric neutrinos, the prompt flux is substantially enriched in the
electron flavor. All of this allowed one to
hypothesize~\cite{Fargion2015prompt-tau, Vissani2019prompt,
Fargion2019prompt} that only the hard, found in $\nu_{\mu}$,
component, have the astrophysical origin, while the one derived from the
analysis of cascade events, rich in $\nu_{e}$ and having the spectrum
closer to $E^{-2.7}$, is explained by prompt atmospheric neutrinos.
However, for the cascades arriving from above the horizon, one can perform
an additional analysis. The atmospheric neutrinos are not born alone -- in
the same atmospheric shower, muons are also produced, which can also be
detected. The requirement of the lack of simultaneous detection of such
muons suppresses the contribution of atmospheric neutrinos coming from
above, including prompt ones, and the distribution of the events selected
in this way becomes anisotropic. Figs.~\ref{fig:atm-prompt-E},
\ref{fig:atm-prompt-ZA}
show the best fit of the observed energy and zenith-angle distributions for
the HESE sample assuming a zero astrophysical component -- an arbitrary
normalization of the prompt neutrino flux is allowed instead. Comparison
with Figs.~\ref{fig:atm-astro-E}, \ref{fig:atm-astro-ZA} indicates that
the spectrum is described by prompt neutrinos equally well as by
astrophysical neutrinos, but  the zenith-angle distribution excludes this
explanation at $>5\sigma$ confidence level. It is worth noting that for a
satisfactory explanation of the spectrum, Fig.~\ref{fig:atm-prompt-E}, the
prompt flux normalization was increased by a factor of 21.56 as compared
to the predictions of modern theoretical models.

It is pertinent to add that an important criterion for the presence of
astrophysical neutrinos is~\cite{Capone, Fargion2015prompt-tau,
Vissani2018tau} the presence of $\nu_{\tau}$ among observed events, since
for atmospheric neutrinos, both conventional and prompt, the fraction
of $\nu_{\tau}$ is very small. Observations of the first candidates for
$\nu_{\tau}$ events \cite{IceCube2bangObs} support the astrophysical
explanation, but the statistics here are still  very small.

\begin{figure}
\centerline{\includegraphics[width=\columnwidth]{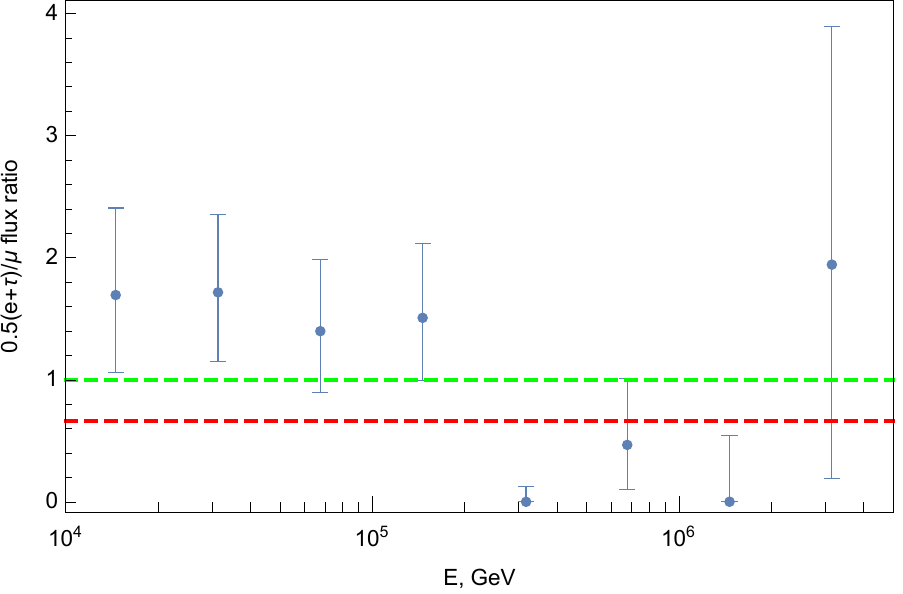}}
\caption{
\label{fig:ratio}
Ratio of neutrino fluxes of different flavors as a
function of energy. Average $\nu_{e}$ and $\nu_{\tau}$ fluxes are taken
from the experimental points of the
analysis \cite{Cascades2020e-tau, HalzenTalk}, $\nu_{\mu}$ -- from the
power-law fit \cite{IceCube-mu2019}. The green dashed line is unity,
predicted in normal $\pi$-meson decays, the red dashed line is 2/3,
predicted in the muon damp mode (see Sec.~\ref{sec:general:pi-mesons}). }
\end{figure}

\paragraph{Flavor composition?}
\label{sec:exper:spec-flav:flav}
The spectra (\ref{Eq:plaw}) with parameters from Table~\ref{tab:plaw-fits}
were obtained under the assumption of equality of the fluxes of
neutrinos of the three flavors. Since
in the analyses based on muon tracks, the atmospheric background is higher
due to the presence of both atmospheric neutrinos and atmospheric muons,
the spectra obtained from tracks are dominated by higher
energies. Such systematics can lead to a dependence of the reconstructed
energy on the flux, if in fact this assumption is violated, cf,
e.g.~\cite{2comp-Chen}. In Fig.~\ref{fig:ratio},
the ratio
of the $\nu_{e}$ and $\nu_{\tau}$ flux to that of $\nu_{\mu}$,
calculated  in energy bins~\cite{Cascades2020e-tau, HalzenTalk}
 from the power-law fit~\cite{IceCube-mu2019} in the energy range common
to the two analyses, is given (deviations of the real spectrum from
the power law, discussed below, would lead to a change in this
picture). It can be seen that at energies $\gtrsim 200$~TeV the results
are in a poor agreement with the standard assumption of equality between
the flavors; the agreement does improve if one assumes a transition to a
different flavor ratio in this energe range. We will see
in Sec.~\ref{sec:general:pi-mesons} that this may indeed happen for sources
with very strong magnetic fields -- so strong that it is not easy,
though in principle possible, to reconcile the model with other
observational data.

At present, the flavor composition of high-energy neutrinos
is not very precisely defined, and available results and perspectives
are discussed in Refs.~\cite{BustamanteFlavour2019,
Flavour2020future}.

\paragraph{North-South Anisotropy?}
\label{sec:exper:spec-flav:aniso}
Another assumption used in the estimation of the spectra is the
isotropy of the diffuse flux. However, none of the approaches to finding
astrophysical neutrinos guarantees a uniform sensitivity to
the flux coming from different directions. This anisotropy in addition
varies with energy, as well as from one analysis to another. Among
early attempts to explain the difference in the spectra recovered from
cascade and track IceCube events, there were also assumptions about
global anisotropy of neutrino arrival directions, see e.g.
Ref.~\cite{Vincent:North-South}. Indeed, the track analysis is more
sensitive to neutrinos arriving ``from below,'' and at the highest
energies -- close to the horizon; the cascade analysis covers the
entire sky and is sensitive to directions ``from above.'' The presence of
a significant anisotropy of the astrophysical flux, e.g, associated with
the dominance of a nearby source in the Southern hemisphere, could explain
the observed discrepancy in the spectra. To date, no such
anisotropy has been detected. The Northern detectors, Baikal-GVD and
KM3NeT, could definitively rule out or confirm such an explanation, since
for them the notions of ``below'' and ``above'' refer to other parts of the
sky.

\paragraph{Two components of the spectrum?}
\label{sec:exper:spec-flav:2comp}
In light of what has been said, it would be natural to simply believe that
the spectrum over the entire investigated interval of energies
differs from the power law, see  Fig.~\ref{fig:tyes}, and the flux
at energies of tens of TeV is higher than expected from the
extrapolation of the power law valid above $\sim
200$~TeV downward in energies. In Sec.~\ref{sec:general}, we will see that
it is extremely difficult to describe
theoretically all the observational data if one assumes the origin of
\textit{all} neutrinos with TeV to PeV energies in sources of the same
type. The natural next step is to assume two components of the spectrum,
one of which (probably extragalactic) has a hard spectrum and extends far
into the PeV range, and the other (possibly related to our Galaxy)
dominates below $\sim (100\dots200)$~TeV~\cite{2comp-Chen, 2comp-Vissani,
2comp-Neronov, 2comp-Vissani-2}. Below we discuss in more detail the
observational motivation for such a hypothesis and possible scenarios for
the origin of the two components of the diffuse neutrino flux.

\subsection{Arrival directions}
\label{sec:exper:aniso}
To a large extent, the methods for analyzing arrival directions were
inherited by high-energy neutrino astronomy from
studies of cosmic rays of ultra-high, $\gtrsim 10^{18}$~eV,
energies (see e.g, \cite{ST-UFN-CR}), whose sources are also unknown.
These methods include searches for and constraints on large-scale (of the
scale of the entire sky) anisotropy, in particular of that related to the
Galaxy's disk or halo, and for deviations from the random distribution of
arrival directions at angular distances comparable to the telescope's
resolution -- namely, the search for autocorrelations, indicating the
existence of point sources, and correlations with specific populations of
theoretically motivated sources. Unlike for cosmic rays, whose trajectories
are deflected by magnetic fields that are often unknown, for neutrinos
it is less important to search for medium-scale ``spots'' on the
celestial sphere, but a new possibility of spatiotemporal
correlation with flaring sources becomes available.

\begin{figure*}
\centerline{\includegraphics[width=\textwidth]{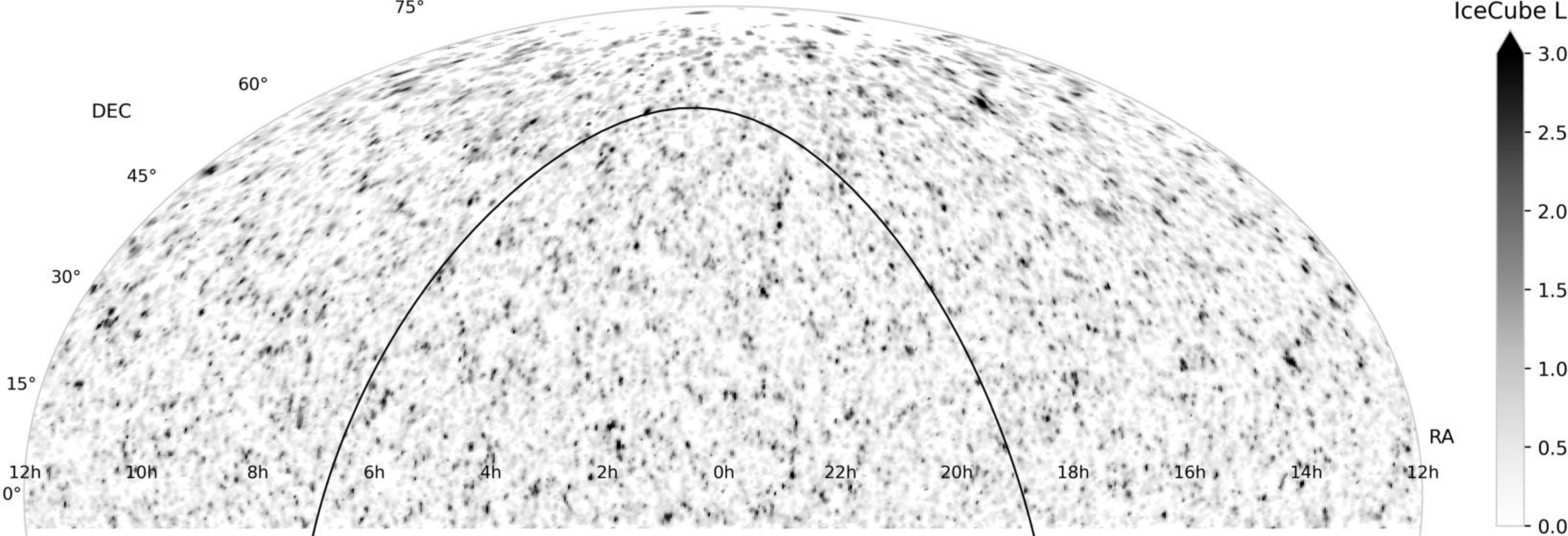}}
\caption{\label{fig:skymapL}
Weighted distribution of IceCube track-event arrival directions
from below the horizon over 7 years of operation (based on
\cite{IceCube7yrSources, IceCube7yrData}). The likelyhood function
describing the probability of detecting a source of astrophysical
neutrinos in a given direction, taking into account the number of events,
the accuracy of determining their arrival directions and the values of the
reconstructed energies, is presented. Equatorial coordinates; the Galactic
plane is shown by a solid black line. The author is grateful to A.~Plavin
for his help in preparing this plot.}
\end{figure*}

\begin{figure}
\centerline{\includegraphics[width=\columnwidth]{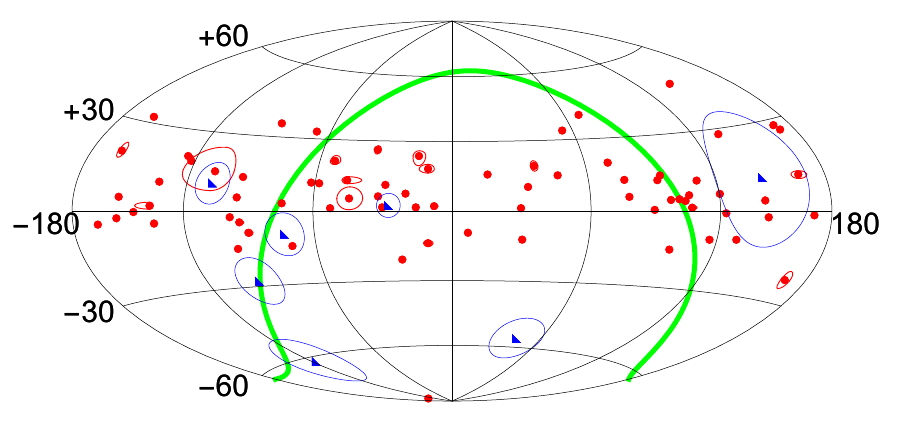}}
\caption{\label{fig:skymapH}
Distribution of IceCube event arrival directions with published
energies above 200~TeV,
Refs.~\cite{IceCube-mu2016, IceCubeEHEAcatalog, IceCube-HESE1,
IceCube-HESE2, ps1710.01179p54, HESE2020, ps1710.01179p31} and GCN and
AMON online alerts. The tracks are shown by dots, the cascades are shown
by triangles, lines indicate areas of 90\% statistical uncertainty of
arrival directions (not always distinguishable for tracks). About half of
these events are atmospheric. Equatorial coordinates; the Galactic plane
is shown as a solid bold line.
}
\end{figure}

A general impression of the distribution of neutrino event arrival
directions is given by the maps of the sky shown in
Fig.~\ref{fig:skymapL}
(all events, i.e., mostly low energies), and Fig.~\ref{fig:skymapH}
(highest energies, largest fraction of astrophysical neutrinos). In both
cases, no obvious differences from the distribution expected for an isotropic
incident flux are detectable. Nevertheless, more subtle analyses, to
discussing of which we turn now, allow one to identify or
constrain the contributions of various astrophysical sources.

\subsubsection{Constraints on the Galactic anisotropy}
\label{sec:exper:aniso:Gal}
Neutrinos of extragalactic origin are collected from all over the Universe,
so their flux must be isotropic to a high degree of accuracy. This
distinguishes them from, for example, ultrahigh energy cosmic rays,
whose mean free path length is
hundreds of megaparsecs, so that
their arrival directions could indicate a heterogeneous distribution
of sources in that volume (for neutrinos, quantitative estimates of such
effect are presented e.g.\ in Ref.~\cite{TamborraMultiplets2MRS}).
If, however, a significant fraction of the observed neutrinos are produced
in our Galaxy, the inhomogeneities in the source distribution should show
up in the anisotropy of the arrival directions. Unfortunately, with a
significant background of atmospheric events it is difficult to search for
such anisotropy; some interesting results have been nevertheless obtained.

\begin{figure}
\centerline{\includegraphics[width=\columnwidth]{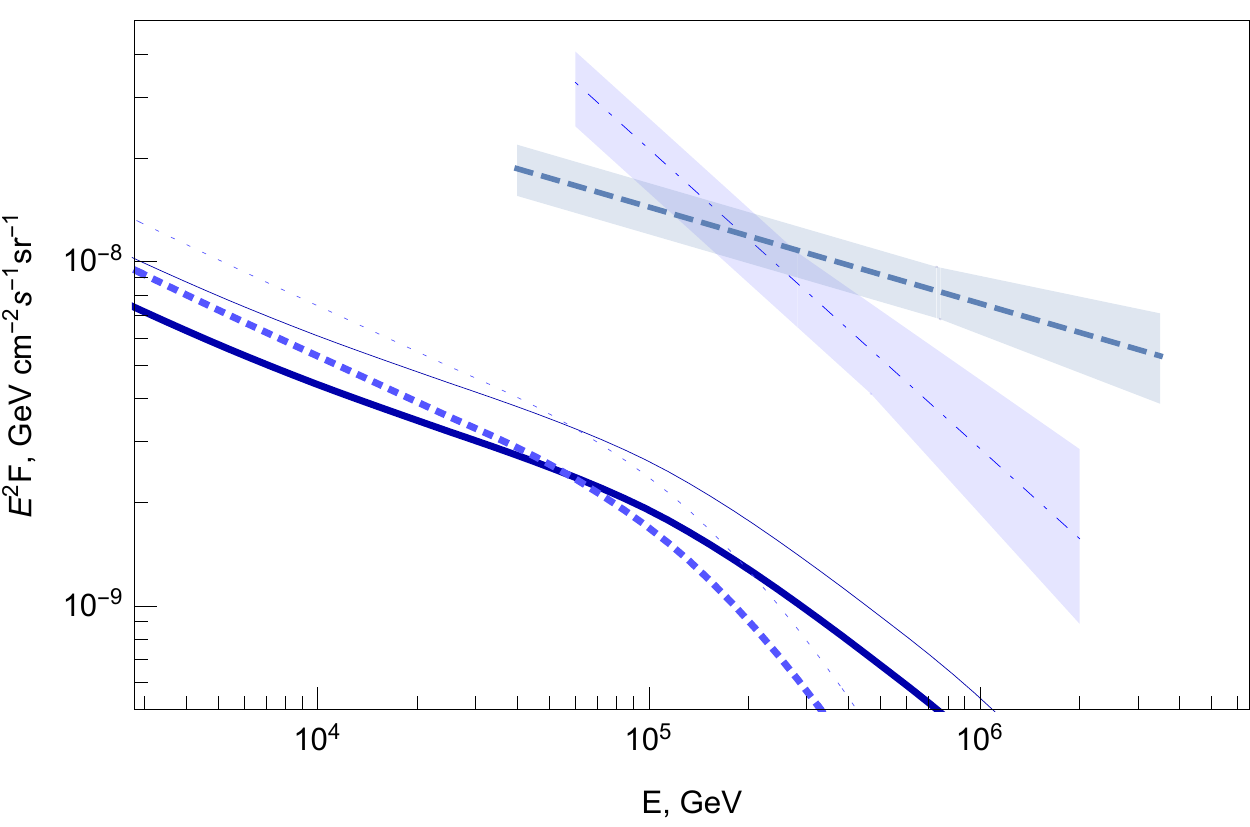}}
\caption{\label{fig:KRAgamma}
Neutrinos from the Galactic plane assuming proportionality
of the flux to the predictions of the KRA$\gamma$ model (in terms of
one-flavor flux $\nu+\bar\nu$). Bold curves -- the best fit from
IceCube cascade events \cite{IceCube7yrCascades}, thin curves -- the
strongest upper limits (90\% CL) from the joint analysis of
IceCube and ANTARES~\cite{IceCubeANTARES-GalPlane} (solid lines --
KRA$\gamma$ with cosmic-ray cutoff at 50~PeV, dotted -- at 5~PeV). The
lines with the shaded uncertainty region show the total astrophysical
neutrino fluxes from IceCube data (blue dashed line -- $\nu_{\mu}$
2019~\cite{IceCube-mu2019}; thin blue dot-dashed line -- HESE
2020~\cite{HESE2020}). }
\end{figure}

\paragraph{The Galactic plane.}
The distribution of visible matter in the Galaxy is very different from
isotropic -- stars and gas are mostly contained in the disk visible
in the sky as the Milky Way. For a quantitative answer to the question of
what fraction of neutrinos may be associated with sources in the disk, the
most effective way is to look for ``the Milky Way in neutrinos'' against a
background of uniformly (up to the experimental exposure) distributed
atmospheric and extragalactic neutrinos. However, an ambiguity arises
here, because the specific distribution of neutrino arrival directions
from the disk depends on the assumption about the sources. Often, one uses
the models constructed to explain the observed Fermi LAT direction- and
energy-dependent intensity of the Galactic diffuse gamma rays. Their free
parameters include not only the distribution of matter and magnetic field
in the Galaxy, but also the characteristics of the cosmic-ray flux. In
particular, the spectrum
and composition of cosmic particles, measured in the Solar system,
are extrapolated to the entire Galactic disk. Obviously,
models of this type should be considered as estimates, and the scatter of
their predictions for the neutrino flux demonstrates this systematic
uncertainty. Recent observational results rely on the model of
Ref.~\cite{KRAgamma}, called KRA$\gamma$, and consider two variants of it
which assume the cosmic-ray spectrum cutoff either
at 5~PeV or at 50~PeV (note that, in general, the results of
cosmic-ray studies indicate their probable Galactic origin at higher
energies as well). In 2019, from the analysis of the anisotropy of IceCube
cascade events \cite{IceCube7yrCascades}, an indication was obtained
(with the statistical significance of $2\sigma$) to the presence of
the Galactic component of the neutrino flux with the shape of the spectrum
and the spatial distribution described by the KRA$\gamma$ model, with the
best-fit normalization of 0.85 (0.65) from the model prediction for
cutoffs of 5 (50) PeV, respectively. The most stringent upper limits (the
coefficient of normalization $<1.19$ for the 5~PeV cutoff and $<0.90$
for the 50~PeV, 90\% CL) were obtained from the joint analysis of IceCube
and ANTARES~\cite{IceCubeANTARES-GalPlane}. As it can be seen from
Fig.~\ref{fig:KRAgamma},
if the detection of this component is confirmed, it would amount to $\sim
10\%$ of the IceCube astrophysical flux,
in rough agreement with theoretical estimates. However,
the mechanism of neutrino production in
interactions of cosmic rays with the interstellar medium
could be not unique
for the Galactic disk; in particular, individual sources accelerating
cosmic rays are present there. This analysis does not allow one to
constrain their contribution. In this context, a nonparametric comparison
of the observed and isotropic distributions of neutrinos in the Galactic
latitude looks more universal. It allows one to find or constrain any flux
component associated with the Galactic disk (see e.g.,
\cite{Neronov-GalPlane, ST-Gal}).

\paragraph{Galactic center -- anticenter.}
Another class of models, including those associated with dark-matter decays
or annihilation and with the interactions of cosmic rays in
the circumgalactic gas (see Sec.~\ref{sec:gal} below), predicts a
different kind of Galactic anisotropy. In this case, the sources
are not concentrated in the disk, but are distributed in the Galaxy in a
more or less spherically symmetric way, with concentration decreasing
from the Galactic center down to the virial radius,
$\gtrsim 200$~kpc. Together with the non-central position of the Sun in
the Galaxy, this leads to the dipole-like anisotropy of arrival directions
with the maximum of the flux from the center of
the Galaxy and the minimum from the opposite
direction~\cite{DubovskyGalDipole}. It is also convenient to search for
such anisotropy, sensitive to the details of the source distribution, in
the data with nonparametric tests~\cite{ST-Gal}. Recently, experimental
data analyses have used this approach in the context of particular
dark-matter models~\cite{DM-1907.11193, DM-1910.12917} together with the
analysis of the contribution of dark matter to the observed neutrino
spectrum~\cite{DM-1804.03848, DM-1903.12623}. Increasing the sensitivity
to particular models, these studies inevitably lose in universality. In
most cases, adding a contribution with such anisotropy leads to an
improvement in the description of the data, but this improvement is not
statistically significant.

\begin{figure}
\centerline{\includegraphics[width=\columnwidth]{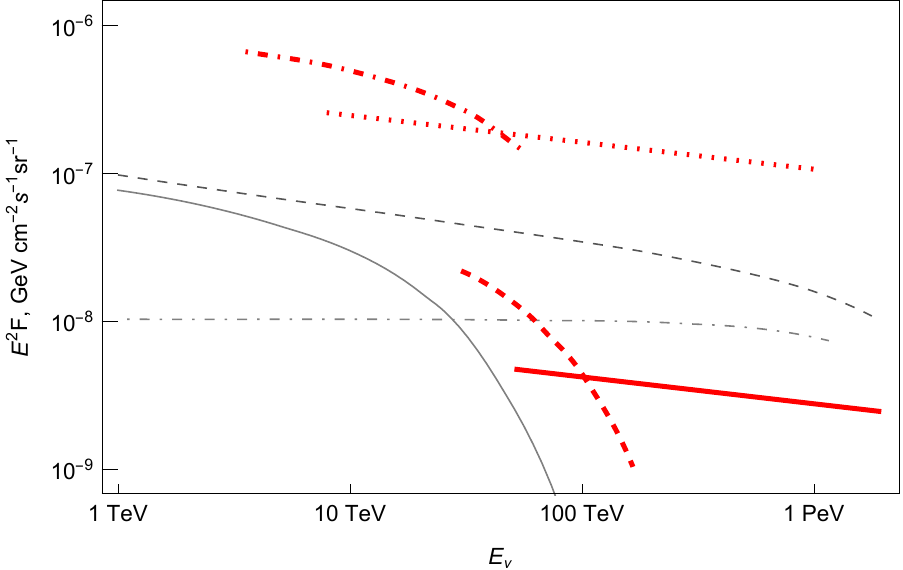}}
\caption{\label{fig:FermiBubbles}
Constraints on models of neutrinos from Fermi bubbles. Bold red --
upper limits on the neutrino flux ($\nu+\bar\nu$, one flavor, under the
$E^{-2.18}$ spectrum assumption) from ANTARES
\cite{ANTARES-FermiBubbles} (dot-dashed line -- spectral cutoff at
50~TeV, dashed line -- no cutoff) and IceCube~\cite{IceCube7yrSources}
(dashed line -- spectral cutoff at
50~TeV, solid line -- no cutoff).
Thin gray lines --  proposed models (dashed line --
\cite{LunardiniBubbles}, dot-dashed line --
leptohadronic~\cite{FangBubbles} and solid --
hadronic~\cite{FangBubbles}). }
\end{figure}

\paragraph{Large Galactic structures.}
The distribution of gas in the Galaxy and in the circumgalactic space is
heterogeneous. In terms of the models
of the neutrino origin discussed in the literature, one singles out
so-called Fermi bubbles, large-scale formations above and below the
Galactic disk in its central region~\cite{FermiBubbles},
possibly related to the past activity of the Galactic nucleus. Their
contribution to neutrino fluxes can be constrained by treating them as
extended sources, to which a higher flux of diffuse radiation in
the sky map corresponds. The results of such analyses did not reveal any
excess of neutrinos from the Fermi bubbles and allowed one to rule out some
models, see Fig.~\ref{fig:FermiBubbles}.

A particular prediction of anisotropy arises also for the model of
the Local Bubble (see Sec.~\ref{sec:gal}), inside which the Solar system
is located. The arrival directions of neutrino events in this model should
concentrate in spots whose positions are determined by the direction of
the Galactic magnetic field lines in the vicinity of the Sun. Because of
great uncertainty in the knowledge of the Galactic magnetic field and of
difficulties in quantitative analysis, the search for such anisotropy
remains an interesting task for future.

\subsubsection{``Blind'' search for point sources}
\label{sec:exper:aniso:blindsearch}
Given the considerable uncertainty regarding theoretically motivated
high-energy neutrino sources, the so-called blind search for their possible
sources is becoming popular. Within this approach, one searches the
entire sky for clusters of neutrinos coming from close directions that
could indicate the most powerful point sources. The modern version
implements this by constructing the maximum likelihood function, which
takes into account not only the arrival directions, but also the accuracy
of their determination, which differs from event to event, as well as the
energies: a cluster of even a few neutrino events allows one to estimate
the hypothetical
neutrino spectrum from a given direction, and if it is harder than the
atmospheric one, then the probability that we are dealing with a real
source and not with a background fluctuation is higher. Based on such maps,
of which Fig.~\ref{fig:skymapL} is an example, many of the results of
this and the next subsections have been obtained.

A significant disadvantage of the blind approach is that only the most
powerful sources could be revealed in this way with high statistical
significance. Indeed, in analyzing each individual direction, it is
possible to determine the statistical probability that the cluster of
events with all its characteristics arises as a result of a random
fluctuation. To do this, it is sufficient to run a significant number of
simulations of random sets of events and to calculate how often the
observed, or larger, value of the likelihood function is achieved due
to fluctuations in these random sets. This probability\footnote{For
convenience, it is often recalculated into standard deviations
for a normal distribution and is said to be significant at a certain
amount of $\sigma$. Although we will also follow this tradition in this
article, it is worth remembering that only the probability value itself is
meaningful, since all involved distributions tend to be non-Gaussian.} is
the main characteristic of the significance of finding a hypothetical
source. However, here we should distinguish between the probabilities of
accidentally finding a ``source'' (1)~in a given predetermined direction
and (2)~anywhere in the sky. In the blind source search
problem, we are dealing with the case (2). The corresponding probability is
much larger, which means that even a pronounced cluster of arrival
directions can easily turn out to be random. If the probability of a
random match for a given direction (pre-trial) is $p_{1}$, then for $N$
independent attempts, the probability (post-trial) becomes $p\sim N
p_{1}$. If the assumption of independence of attempts is violated, $p$
is determined from a Monte-Carlo simulation. The ratio $p/p_{1}$, called
the penalty factor (see e.g.,~\cite{CutsAndPenalties}), for a blind search
of neutrino sources is easy to estimate approximately. Assume that the
area of the uncertainty of the arrival direction of a track event is of
the order of 2 square degrees, and half of the celestial sphere, that is,
$2\pi$ steradian, is involved in the analysis. Then we find the number of
quasi-independent directions as the ratio of these areas, $\sim 10^{4}$ (a
more correct estimate is required to take into account the anisotropy of
the exposure). This means that to assert a $3\sigma$ statistical
significance of a source found in a blind search (i.e., $p\sim 10^{-3}$),
one would need to have $p_1\sim 10^{-7}$, that is, the pre-trial
significance $>5\sigma$.

The most significant blind-search result for the dataset with the largest
exposure, 10 years of IceCube track events, corresponds to
the direction in the sky, which is $0.35^{\circ}$ away from the galaxy
NGC~1068, a nearby galaxy with intense star formation. For
this direction, $p_{1}=3.5 \times 10^{-7}$, $p=9.9 \times 10^{-2}$,
which corresponds to the post-trial significance of $\sim 1.6\sigma$.
Additionally it should be noted that the fit indicates a very soft
neutrino spectrum from this direction: the power-law index of the
spectrum is 3.4, consistent with that expected for atmospheric,
rather than astrophysical, neutrinos. Thus, at the time of writing this
review, no individual source in the sky has been detected in a blind
search with post-trial significance exceeding even $2\sigma$.

A way to increase the sensitivity of such an analysis to weaker
sources is to use a priori fixed catalogs
of theoretically motivated neutrino sources. If the sources in the
catalog are of one and the same type, then we are talking about a
population analysis, and the cumulative signal from many, even weak,
sources can be registered by summing multiple signals from their
directions, each of which is statistically insignificant. This will be
discussed in Sec.~\ref{sec:exper:aniso:stacking}. Here we mention
recent attempts to increase the sensitivity of the blind search by
combining different types of sources. A catalog of specific sources whose
neutrino origins have been discussed in the literature is compiled and
supplemented by the sources somewhat similar to them. A wide variety of
objects fall into the same list. A blind search is limited to the
directions of these ``motivated'' sources, and the penalty factor
becomes equal to the number of sources in the catalog, i.e. instead of
$\sim 10^{4}$ it becomes, for example, $\sim 10^{2}$. The danger of this
approach is that some of the sources were actually ``motivated'' by
previous, based on partial neutrino samples, analyses, which revealed
excesses of neutrino events from their directions. Such an analysis can
only make astrophysical sense if along with the answer to the question,
from which sources the excesses of neutrino events are seen, compared to
random, there will also be an answer why they are not seen from other
similar sources.

Nevertheless, even along this path, no
single source with a post-trial significance $>3\sigma$ was found.
Ref.~\cite{IceCube10yrSources} analyzed a catalog of 110
gamma-ray sources identified from observational data in different bands,
based on various criteria. This included 12 sources in the Galaxy, 3
nearby galaxies -- the Large and Small Magellanic Clouds and the
Andromeda galaxy, M31, 4 starburst galaxies, 89
active galactic nuclei and 2 unidentified sources. All sources were
studied as point sources, although the angular sizes of some Galactic
sources, and especially of nearby galaxies, are noticeably larger than the
IceCube angular resolution for tracks. For two sources the post-trial
significance was $>2\sigma$, they are the already mentioned galaxy
NGC~1068 and the blazar TXS~0506$+$056, which will be discussed in more
detail in Sec.~\ref{sec:exper:aniso:flares}. Additionally, in
Ref.~\cite{IceCube10yrSources} it has been observed that another two other
sources (blazars PKS~1424$+$240 and GB6~1542$+$6129) have pre-trial
$p_{1}<2\times 10^{-3}$ and that the presence of 4 sources out of $\sim
100$ with such low $p_{1}$ is itself unlikely for
a random fluctuation. The statistical analysis indicated a
probability of such a of coincidence $\sim 5\times 10^{-4}$ (significance
$\sim 3.3\sigma$). This result, however, does not take into account the
aforementioned arbitrariness in the compilation of the source catalog; in
particular, the removal from it of TXS~0506$+$056, widely discussed
previously, reduces the significance to
$2.3\sigma$~\cite{IceCube10yrSources}. The authors of
Ref.~\cite{IceCube10yrSources} conclude from this analysis that the data
indicate the presence of real sources in their catalog; however an
astrophysical interpretation of this result is difficult.

So far we have mainly referred to the results of the search for neutrino
sources in the Northern sky from IceCube track events. Similar work has
been done for the Southern sky in the joint analysis of IceCube and
ANTARES~\cite{ANTARESandIceCubeSources} and also did not result in the
detection of statistically significant individual sources, including
those in the region of the Galactic Center.

The lack of manifestations of significant individual sources in the
analysis described in this section indicates that the observed flux of
astrophysical neutrinos is produced by a large number of not very strong
sources. This fact can be used to constrain the source models, see
Sec.~\ref{sec:general:aniso} below.

\subsubsection{Source populations}
\label{sec:exper:aniso:stacking}
The next natural step after a blind search is to take a population
of the same type of astrophysical objects that are potential
neutrino sources, and check whether the entire population
has an excess of arrival directions compared to a random
distribution. In this way one can find or constrain the coherent
effect from many sources of the same class, each of which is not strong
enough to be detected in a statistically significant way in an
individual or blind analysis. Technically, such an analysis can be
performed either using the likelihood function described in
Sec.~\ref{sec:exper:aniso:blindsearch} (test statistics are based on the
sum of the values of the likelihood function for all directions from the
sources in the catalog), or by counting the number of sources from the
catalog in the area near each neutrino arrival direction (in this case,
the test statistics is simply the total number of ``source-neutrino
pairs''). In both cases, a weight can be assigned to each source, e.g.,
depending on the flux of its electromagnetic radiation in a particular
band or from other considerations. In the second approach, the size and
the shape of the region are given by the accuracy of determination of the
neutrino arrival direction (e.g., at the 90\% CL). These approaches have
their pros and cons: the likelihood function allows one to account for
10\% of events that fall outside the 90\% CL angular resolution, and may
contain additional information, encoded in continuous characteristics of
events, such as energy, influencing the probability that a given neutrino
is astrophysical. At the same time, when counting events, it is easier to
account for the poorly known systematic error in determination of the
arrival direction of neutrino, discussed in
Sec.~\ref{sec:intro:detection} (see e.g., \cite{neutradio1,
ps2001.09355Resconi}). Typically, the likelihood function is used
for large numbers of neutrino events (low energies) or for extended
uncertainty regions of complex shape (cascades).

Table \ref{tab:popu1} provides a summary
of numerous (perhaps not all) published analyses of correlations
of neutrino arrival directions with potential source populations.
\begin{table*}
\begin{center}
\begin{tabular}{cccccrlc}
\hline
\hline
Source & Number & Neutrino & Method & Reference &
\multicolumn{2}{c}{Post-trial} & Contribution \\
sample & of sources & sample & & & \multicolumn{2}{c}{significance} & to
the flux\\
\hline
\multicolumn{8}{c}{Galactic sources}\\
\hline
Supernova remnants & 23 & I $\nu_{\mu}$ (10)& L &
\cite{IceCube10yrSources} & \multicolumn{2}{c}{--} &\\
Pulsar wind nebulae & 33 & I $\nu_{\mu}$ (10)& L &
\cite{IceCube10yrSources} & \multicolumn{2}{c}{--} &\\
Pulsar wind nebulae & 35 & I $\nu_{\mu}$ (9.5)& L &
\cite{ps2003.12071} & \multicolumn{2}{c}{--} &$<48\%$~$\nu_{\mu}$\\
Unidentified gamma-ray& \multirow{2}{*}{23} & \multirow{2}{*}{I $\nu_{\mu}$ (10)}&
\multirow{2}{*}{L} &
\multirow{2}{*}{\cite{IceCube10yrSources}} & \multicolumn{2}{c}{\multirow{2}{*}{--}} &\\
 sources $>100$~GeV &&&&&&\\
\hline
\multicolumn{7}{c}{Extragalactic sources other than blazars}\\
\hline
Starburst galaxies& 45 & I $\nu_{\mu}$ (3)& L &
\cite{ps2007.12706}
& \multicolumn{2}{c}{--} &\\
Starburst galaxies& 64 & A (10)& L &
\cite{ps2012.15082}
& \multicolumn{2}{c}{--} &\\
Radio galaxies& 63 & A (10)& L &
\cite{ps2012.15082}
& \multicolumn{2}{c}{--} &\\
Dust-obscured galaxies& 15 & A (10)& L &
\cite{ps2012.15082}
& \multicolumn{2}{c}{--} &\\
Gamma-bright (4LAC)&\multirow{2}{*}{65} & \multirow{2}{*}{I $\nu_{\mu}$ (3)}
& \multirow{2}{*}{L} &
\multirow{2}{*}{\cite{ps2007.12706}}
& \multicolumn{2}{c}{\multirow{2}{*}{--}} &\\
AGN other than blazars &&&&&&&\\
Large-scale extra- & 94 & I HE all& C &
\cite{ps1804.01386}& \multicolumn{2}{c}{--} &\\
galactic jets &&&&&&&\\
\hline
\hline
\multicolumn{8}{c}{Gamma-selected blazars}\\
\hline
3FGL& 729 & I $\nu_{\mu}$ $200+$& C &
\cite{ps1611.06338}& \multicolumn{2}{c}{--} &\\
2FHL BL Lacs&& I $\nu_{\mu}$ $200+$& C&
\cite{ps1702.08779}& \multicolumn{2}{c}{--} &\\
2FHL HBL& 149 & I $\nu_{\mu}$ (7)& L &
\cite{ps1710.01179p31}& \multicolumn{2}{c}{--} &$<27\%$~$\nu_{\mu}$\\
2FHL HBL& 149& I HE all& C&
\cite{ps1601.06550Resconi}& \multicolumn{2}{c}{--} &\\
3FHL & 1301 & I $\nu_{\mu}$ (8)& L &
\cite{ps1908.08458}& \multicolumn{2}{c}{--} &$<17\%$~$\nu_{\mu}$\\
3FHL & 1301 & I HE $\nu_{\mu}$& C &
\cite{ps2001.09355Resconi}& \multicolumn{2}{c}{--} &\\
3LAC HBL & 386 & I HE $\nu_{\mu}$& C &
\cite{ps1601.06550Resconi}& \multicolumn{2}{c}{--} &\\
3LAC & 1255 & A (10)& L &
\cite{ps2012.15082}
&\multicolumn{2}{c}{--} &\\
3LAC FSRQ & 414 & I $\nu_{\mu}$ (7) & L &
\cite{ps1710.01179p31}
&\multicolumn{2}{c}{--} &\\
4LAC & 2796 & I $\nu_{\mu}$ (3) & L &
\cite{ps2007.12706}& \multicolumn{2}{c}{--} &$\lesssim 11\%$
$\nu_{e,\tau}$\\
4LAC & 2794 & I HE $\nu_{\mu}$ & C &
\cite{ps2001.09355Resconi}& \multicolumn{2}{c}{--} &\\
VOU$\gamma$ HBL && I HE $\nu_{\mu}$ & C &
\cite{ps2001.09355Resconi}& \multicolumn{2}{c}{$3.2\sigma$}
&$\sim21\%$~$\nu_{\mu}$\\
\hline
\multicolumn{8}{c}{Blazars selected by other criteria}\\
\hline
2WHSP& 1681 & I $\nu_{\mu}$ (7)& L &
\cite{ps1710.01179p31}& \multicolumn{2}{c}{--} &\\
2WHSP& 1681& I HE all& C &
\cite{ps1601.06550Resconi}& \multicolumn{2}{c}{--} &\\
3HSP & 2011 & I HE $\nu_{\mu}$ & C &
\cite{ps2001.09355Resconi}& \multicolumn{2}{c}{$2.8\sigma$} &\\
RFC & 3388 & I $\nu_{\mu}$ $200+$ & C &
\cite{neutradio1}& $3.1\sigma$
&\multirow{2}{*}{$\left.\!\protect\vphantom{\frac{A^{2^{6}}}{A_{2_{6}}}}\right\}4.1\sigma$}&\\
RFC & 3411 & I $\nu_{\mu}$ (7) & L &
\cite{neutradio2}& $3.0\sigma$ && $\sim25\%$~$\nu_{\mu}$\\
RFC & 3388 & I $\nu_{\mu}$ (10) & L &
\cite{ps2103.12813}& \multicolumn{2}{c}{--} &$\lesssim30\%$~$\nu_{\mu}$\\
\hline
\hline
\end{tabular}
\end{center}
\caption{\label{tab:popu1} Results of searches for correlations of
high-energy neutrino arrival directions
with populations of astrophysical sources.
\textit{Source
samples:} all Galactic sources are selected from observations above
100~GeV; for the selection of other gamma-ray sources, Fermi LAT catalogs
were used: all sources, 3FGL~\cite{3FGL}; sources detected above
50~GeV, 2FHL~\cite{2FHL}, and above 10~GeV, 3FHL~\cite{3FHL}; gamma-ray
bright active galactic nuclei (AGN), 3LAC~\cite{3LAC} and 4LAC~\cite{4LAC},
-- and the VOU-blazars database~\cite{VOU}. Non-gamma-ray related selection
criteria for blazars included a high-frequency ($>10^{15}$~Hz, HBL)
synchrotron peak, 2WHSP~\cite{2WHSP} and 3HSP~\cite{3HSP} catalogs,
and the presence of a compact core visible with very-long baseline radio
interferometers, RFC catalog (http://astrogeo.org/rfc).
\textit{Neutrino samples:} based on the data from IceCube (I) or ANTARES
(A), $\nu_{\mu}$ -- only track events, number in parentheses -- number of
years of exposure, HE -- published individual high energy events, $200+$
-- energies above 200~TeV. \textit{Method:} L -- likelihood function, C --
coincidence counting. \textit{Post-trial significance:} not specified if
less than $2.5\sigma$. \textit{Contribution to the flux:} estimate of the
fraction of the astrophysical flux assuming a power-law fit for tracks
($\nu_{\mu}$) or cascades ($\nu_{e,\tau}$) associated with the sources of
a given sample.}
\end{table*}
As one can see, there are only a few statistically significant results.
The first fact that draws attention is the lack of
correlations with gamma-ray bright blazars from the Fermi LAT catalogs,
which, before the detailed analysis of IceCube results, were
considered to be among the most likely sources of high-energy neutrinos.
That said, a few statistically significant results point to blasars
selected according to other criteria. Let us briefly discuss them.

By definition, a blazar is a nucleus of an active galaxy that has a
relativistic jet pointing almost exactly at the observer. Relativistic
kinematic effects lead in this case to strong amplification of the
observed radiation flux. The same mechanism of amplification works for
neutrinos. Physical conditions in nuclei and jets allow
protons to be accelerated to the energies required for neutrino production.
All of this together makes blazars very attractive candidates
to high-energy neutrino sources.

While identification
as a blazar is determined only by this
geometric characteristic of the jet, physical conditions in these
sources may be different. They are manifested primarily in
in the bands where the blazar emits the dominant part of its non-thermal
radiation. The main characteristic here is the frequency,
corresponding to the maximum of the synchrotron peak in the broadband
spectral energy distribution, which varies from $\nu_{\rm
peak}<10^{12}$~Hz (radio band -- radio quasars with a flat spectrum, FSRQ)
to $\nu_{\rm peak}>10^{17}$~Hz (X-ray band -- extreme BL Lac type objects,
HBL). In many studies, the primary focus has been on blazars bright in
the gamma-ray band, observed by Fermi LAT (1--100 GeV). These are the
highest energies at which fairly uniform over the sky samples of
sources are available; in addition, at even higher energies, the Universe
becomes not completely transparent to gamma rays because of
$e^{+}e^{-}$ pair production, see Sec.~\ref{sec:general:pi-mesons}.
Unfortunately, due to the insufficient sensitivity of Fermi LAT and
the strong variability of blasars in the gamma-ray band, the samples of
gamma-ray bright sources capture only a small fraction of blazars.
It is
also possible that this fraction is unrepresentative
because the origin of gamma rays in different classes of blazars can be
very different. The universal criterion for the presence of a relativistic
jet directed towards us is the observation of a compact (no more than a few
parsecs in size) region of intense radiation at the center of the source,
located at a gigaparsec-scale
distance~\cite{ParsecJetsReview,JetsReview}. For this one requires the
angular resolution, achievable in modern astronomy only with very-long
baseline radio interferometry (VLBI)~\cite{VLBIbook}. A catalog of such
compact, VLBI-selected, objects, the Radio Fundamental Catalog
(RFC), was the basis for the search for neutrino arrival-direction
coincidences with blazars in Refs.~\cite{neutradio1, neutradio2}.
Specifically, a complete, limited in the compact-component flux at 8~GHz,
sample of about 3400 blazars isotropically distributed over the celestial
sphere, was used.

Ref.~\cite{neutradio1} analyzed a sample of 56 arrival directions
of published IceCube track events with neutrino energies
$E_{\nu}>200$~TeV. The VLBI flux of complete-sample sources near the
neutrino arrival directions (more precisely, in the 90\% CL regions of
statistical uncertainty of the arrival directions plus the systematic
uncertainty estimated in the analysis) was used as the test statistics.
This average flux turned out to be significantly higher than expected for
the random distribution of arrival directions of neutrino events. After
accounting for the penalty factors associated with variation in the
amount of the systematic error, the statistical significance of the
established relationship between VLBI sources and high-energy neutrinos
was found to be $3.0\sigma$. Additionally, Refs.~\cite{neutradio1,
Hovatta} found the coincidence of neutrino arrival times with
radio flares of these correlated blazars, see
Sec.~\ref{sec:exper:aniso:flares}.

In the spring of 2020, the IceCube collaboration published the values of
the likelyhood function
described in Sec.~\ref{sec:exper:aniso:blindsearch}, constructed from seven
years of track-event observations (this is the one presented in
Fig.~\ref{fig:skymapL}). In Ref.~\cite{neutradio2}, based on these
data (for directions from below the horizon, where the flux of atmospheric
muons is strongly suppressed),
the correlation of directions with the maximum probability of locating a
point source of astrophysical neutrinos with directions to blazars from the
same RFC catalog was established
(significance $3.1\sigma$ post
trial). The joint statistical significance of the
observations \cite{neutradio1} and \cite{neutradio2} was estimated as they
were based on the same sample of sources. To make the two analyses
completely independent,  directions
of neutrinos with
energies above 200~TeV,
used in the analysis~\cite{neutradio1} and
recorded during the time period for which the
likelyhood map was constructed,
were cut from the map. The
cumulative statistical significance of neutrino correlations with the
VLBI-selected blazars was found to be $4.1\sigma$. The data underlying
the likelihood function were dominated by events with energies
from a few TeV to several tens of TeV, and the unexpected result of this
analysis was the association of neutrinos of the entire
range of energies studied by IceCube, from TeV to PeV, with sources of the
same class. The sources in the complete sample included in the analysis are
responsible for about 25\% of the astrophysical flux estimated from the
muon tracks (assuming a power law
(\ref{Eq:plaw}) with parameters~\cite{IceCube-mu2019} over the entire
energy range). Taking into account the correction for similar sources
not present in the catalog raises this value to $\sim 100\%$.

It should be noted that in Ref.~\cite{ps2103.12813}, the same set of
VLBI-bright blazars was used for a correlation analysis with the recently
published 10-year IceCube public data set of muon
tracks~\cite{IceCube10yrData, IceCube10yrDataPaper}. Despite the increased,
compared to the 7-year
set~\cite{IceCube7yrSources, IceCube7yrData}  used in
Ref.~\cite{neutradio2}, exposure, this analysis found no significant
correlations. The contribution of blazars from the catalog to the flux of
astrophysical neutrinos was limited to $\lesssim 30\%$, which is consistent
with the result of Ref.~\cite{neutradio2} ($\sim 25\%$). Note that this
IceCube dataset is not well suited to test the hypotheses formulated
earlier, since it contains neither the liklyhood function, nor
reconstructed energies of  individual neutrinos (only muon energies
$E_{\mu}$, see Sec.~\ref{sec:intro:detection} and Fig.~\ref{fig:E-uncert},
are published). The authors of Ref.~\cite{ps2103.12813} constructed their
own likelyhood function, simplified compared to the IceCube function used
in {\cite{neutradio2}, which took into account the neutrino energies. The
sensitivity of such an analysis is low, so it is not surprising that it
neither confirmed nor excluded the result of Ref.~\cite{neutradio2}. The
work to test the hypothesis of a connection of neutrinos with VLBI-bright
blazars continues with the data from Northern hemisphere experiments.
Preliminary results~\cite{ANTARES-RFC} of the ANTARES collaboration
indicate the presence of correlation of arrival directions, compatible
with that found from IceCube~\cite{neutradio2}.

Correlations (significance $3.2\sigma$ post trial) of the arrival
directions of IceCube high-energy track events published as separate
lists and alerts, with blazars selected on the basis of other criteria,
have been found in Ref.~\cite{ps2001.09355Resconi}. The sample of
sources used there was not complete by any criterion, but was a combination
of heterogeneous data based on observations in different bands. The only
requirement was the registration of gamma rays from the blazar by some
instrument. Forty-eight coincidences of directions of arrival neutrinos to
the blazars were found, with an expectation of $\sim 32$ for random
directions. It is interesting to note that 39 of these 48 blazars are
present in the RFC catalog, i.e. that is, they have a VLBI-compact
component, indicating the presence of parsec-scale relativistic jets, and
16 of them belong to the complete sample used in Refs.~\cite{neutradio1,
neutradio2}. Therefore, the effect found in
Ref.~\cite{ps2001.09355Resconi} is actually saturated by sources of
the same class.

Thus, blazars, that is, active galaxies with relativistic
jets pointing almost exactly toward the observer (but not necessarily bright
in the gamma-ray band investigated by Fermi LAT), given the contribution of
sources not listed in the catalogs, can explain the entire astrophysical
neutrino flux, as estimated from the IceCube muon tracks, when
extrapolated down to energies of the order of TeV by a power law
(\ref{Eq:plaw}). In the context of the  two-component model
discussed in Sec.~\ref{sec:exper:spec-flav}, this corresponds to  100\%
of the hard flux component.

\subsubsection{Search for flares}
\label{sec:exper:aniso:flares}
A blind search for neutrino flares from arbitrary directions in the sky
opens up even more freedom in the parameters -- moment and duration of
the flare are added to the direction in the celestial sphere.
The probability of finding a spatio-temporal cluster of events
resulting from random fluctuations is very high, and studies of this kind
in the presence of a large random background make sense only for
predetermined sources, whose variability in the electromagnetic
radiation could be associated with neutrino bursts. The most
highly variable objects among potential neutrino sources are blazars and
one-time catastrophic events on the stellar scales, see e.g,
Ref.~\cite{Murase-transient}. Neutrino flares are often searched in
spatio-temporal coincidence with such events or with flares of
electromagnetic radiation from blazars. Note that often the ``neutrino
flare'' means the moment of arrival of a single rare (high-energy) event
with a high probability of astrophysical origin. However, these results
should be interpreted with
caution: for the Poisson statistics, a single event may signal both an
increase in the average expected rate of arrival of events (a flare), or a
completely random single event at an expected rate of less than one event
per observation time. The estimation of the neutrino flux in this case
is ambiguous~\cite{EddingtonBias}.

\paragraph{AMANDA and 1ES~1959$+$650.}
Historically, the first claim
of a coincidence of  neutrino and gamma-ray flares of a blazar
was made by the AMANDA collaboration based on 2002
data~\cite{Ackerman-thesis, BernardiniAMANDA}. From the direction of the
blazar 1ES~1959$+$650, over 4 years of observations,
5 neutrino events were observed at the expectation of 3.7, which by itself
is consistent with fluctuations. However, 3 of these 5 events came in
a time interval of 66 days, coinciding with  flares of this source in
TeV gamma rays.
The collaboration refrained from assessing the statistical
significance of this coincidence because it was detected a posteriori, not
in a ``blind'' analysis. Theoretical interpretations of this
observation were controversial~\cite{Halzen-AMANDAblazar,
Boettcher-AMANDAblazar}. New similar gamma-ray flares of the same blazar
in 2016 were not accompanied by neutrino excesses at IceCube; this,
however, does not contradict the predictions of the proposed
models~\cite{IceCube-AMANDAblazar}. It was this observation by AMANDA
which motivated \cite{SpieringUFN} the development of a system of mutual
neutrino and gamma-ray telescope alerts, which has been actively developing
already in the IceCube era.

\paragraph{IceCube and TXS~0506$+$056.}
One of the most widely known cases of a coincidence of a detected
IceCube neutrino event with a gamma-ray blazar flare is the
neutrino alert IC170922A, associated with the blazar TXS~0506$+$056.
Unlike its AMANDA predecessors, the IceCube collaboration in this case,
quite similar to the one described above, has evaluated and
published \cite{IceCube-ScienceTXS1} \textit{a posteriori} the statistical
significance of this coincidence. In the description of this well-known
event, we follow the original work~\cite{IceCube-ScienceTXS1}.

A muon track, corresponding with a high probability to an
astrophysical neutrino with energy of hundreds of TeV (an estimate of the
energy of this event is shown in detail in Fig.~\ref{fig:E-uncert}),
was recorded on 22 September 2017. The standard analysis demonstrated a
coincidence of the arrival direction with the position of the gamma-ray
bright blazar TXS~0506$+$056, which was in a period of increased activity.
In response to the IceCube alert, this source was observed by atmospheric
Cherenkov telescopes recording gamma rays with energies $\gtrsim 100$~TeV:
on September 23 -- by HESS and VERITAS and on September 24 by MAGIC. Gamma
rays from this source were not detected by the three instruments; however,
when MAGIC repeated its observation on September 28, it detected a
non-zero signal corresponding to the increase of the flux in comparison
with the upper limits from September 23-24. Multi-wavelength
observations demonstrated that  at the time of the neutrino
arrival, the blazar was at the beginning of a prolonged flare in the radio
band; its X-ray flux was quite modest (this observation has severely
limited theoretical models of the neutrino origin). In the optical band, a
few minutes after the neutrino detection, a strange, short drop in the
brightness of the source was observed~\cite{Lipunov:2020ptp}.

Blasars are numerous and highly variable sources, so the probability
to accidentally find a blazar in a flaring  state in a given
direction and at a given moment in time is quite high. In the absence of
an a priori fixed procedure, a correct estimate of the probability of such
a coincidence is difficult: the main problem is to choose,
what to call ``a flare'' of the source and what to call ``a
coincidence'' with the flare. The authors of Ref.~\cite{IceCube-ScienceTXS1}
proposed and applied several methods for assessing the statistical
significance, described in the Supplementary Information to their paper.
For these estimates, only correlations with Fermi LAT were used, since
only this instrument provides continuous observations of blazars across
the entire sky, to which a particular detected flare could be compared.
The probability of a random coincidence, calculated a posteriori and not
accounting for different trials, was $2.1\times 10^{-5}$ ($4.1\sigma$ pre
trial). The authors of the paper counted observations of other neutrino
events (at the time of the event analysis, there were a total of 51 events
selected using the same criteria as IC170922A), for which there was no
overlap with gamma-ray-bright blazar outbursts, as independent trials.
Taking into account the corresponding penalty factor results in the
probability $\approx 10^{-3}$ ($3.0\sigma$ post trial). This estimate does
not account for the additional penalty factor for testing four models of
the gamma-flux correlation, three of which can be considered independent,
and for the choice of the Fermi-LAT light curve simulation parameters (bin
size in time and the choice of the energy bin). Note also that after the
publication of this result, IceCube has changed the criteria for selecting
alert events, so that a statistically correct analysis
of how many other similar neutrino events were not accompanied by flares
of gamma-ray sources since then is not possible.

From the facts that only for one gamma-ray source such a coincidence has
been reported, and that population analyses do not indicate a significant
association of gamma-ray bright blazars with neutrinos, see
Sec.~\ref{sec:exper:aniso:stacking}, it has been suggested that the blazar
TXS~0506$+$056 is a ``special'' source, so the history of detection of
neutrinos of all energies from this direction was analysed. A flare  of
neutrino events (19 events with an expectation of 6) with energies of (0.1
-- 20)~TeV and a significance of 3.5$\sigma$ (post trial) was
found~\cite{IceCube:TXS-Science2} from that specific direction in the sky.
The time period in which this neutrino flare happened, 2014, was not
special in terms of the gamma-ray activity of the
source~\cite{Halzen-no-gamma-2014-TXS}. Note that using the IceCube event
reconstruction published in 2021
\cite{IceCube10yrData,IceCube10yrDataPaper}, the pre-trial probability of
registering this flare as a result of a random fluctuations is only $8.1
\times 10^{-3}$~\cite{IceCube10yrData}, which is $\sim100$ times higher
than in the original analysis. With the same change in the post-trial
significance it would amount to $\approx2\sigma$.

Numerous papers have investigated multiwavelength characteristics
of this source in order to give a theoretical description
of the neutrino production, see e.g.,~\cite{TXS-1807.04275, TXS-1807.04335,
TXS-1807.04537, TXS-1807.04748, TXS-1808.05540, TXS-1808.05651,
TXS-1812.05654, TXS-1812.05939}. It was not possible to propose a model
that simultaneously describes the events of 2014 and 2017 within a single
mechanism. A study of the characteristics
of the TXS~0506$+$056 source demonstrated that it is a typical
radio blazar~\cite{Kovalev:TXStypical}. Thus, the
observations~\cite{IceCube-ScienceTXS1, IceCube:TXS-Science2} may not
testify to the singularity of this source, but serve to illustrate
the established population connection of neutrinos and radio blazars
\cite{neutradio1, neutradio2}; the gamma-ray flare might be
only a simultaneous signal of the source activity.

\paragraph{Flares of radio blazars.}
This interpretation motivated the search for coincidence of arrival moments
of high-energy IceCube neutrinos with radio flares of the
blazars~\cite{neutradio1} located in the same directions. The result of
this study is illustrated in Fig.~\ref{fig:radioflares},
\begin{figure}
\centerline{\includegraphics[width=\columnwidth]{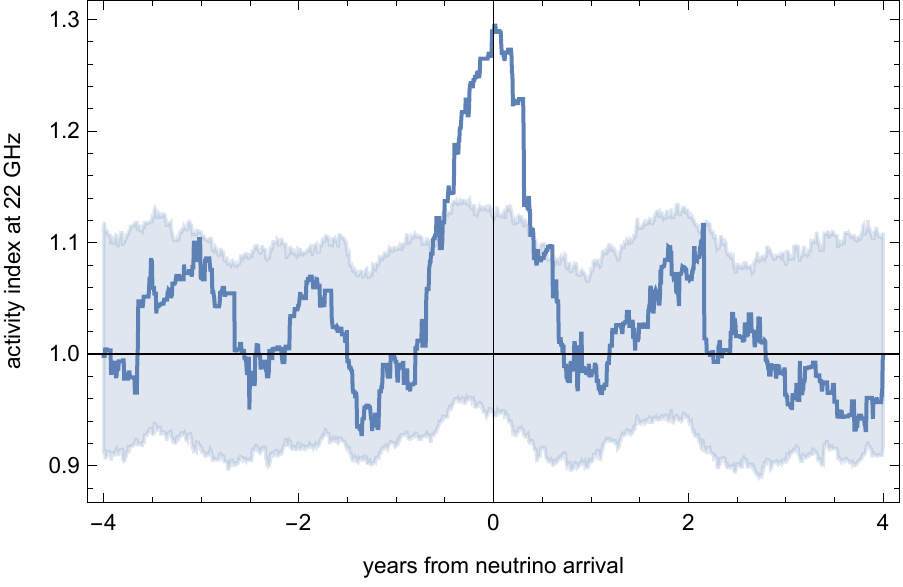}}
\caption{\label{fig:radioflares}
Coincidence of the arrival moments of $E_{\nu}>200$~TeV neutrinos with
flares of blazars selected from the VLBI catalog (22~GHz RATAN-600 data
from Ref.~\cite{neutradio1}). The activity index is defined as the ratio of
the radio flux in the optimized period of 0.9 year to the flux for the
entire period of observations. Averaged over 18 sources coinciding with the
neutrino arrival directions. The shaded area is the expectation for random
arrival directions (68\% CL).}
\end{figure}
from which one can see that, on average, the blazars coinciding with the
neutrino arrival directions are in the period
of the radio flare at the time of the neutrino arrival.
Despite the impressive appearance of the figure, the post-trial
statistical significance of this result is only slightly greater than
$2\sigma$ due to the penalty factor for the selection of the optimal
time-window width. This result, obtained with the data of the blazar
monitoring program on the RATAN-600 telescope (Special Astrophysical
Observatory of the Russian Academy of Sciences), was confirmed by
observations of the American OVRO and Finnish
Metsahovi observatories~\cite{Hovatta}.

\paragraph{Cosmic gamma-ray bursts.}
These stellar-scale processes with extreme
energy release~\cite{Postnov-GRB-UFN, Bykov-GRB-UFN, Aptekar-GRB-UFN}
have long been considered as potential sources of high-energy
protons and neutrinos, see e.g.\ \cite{Milgrom:GRB, Waxman:GRB, Vietri:GRB,
Waxman:GRBnu}, but no space-time correlation
between neutrinos and detected gamma-ray bursts has been
found~\cite{IceCube:2016GRB, IceCube:2017GRB, ANTARES:2020GRB,
ANTARES:VHE-GRB}. This narrows the range of gamma-ray bursts -- potential
sources of high-energy neutrinos -- to optically thick ones not
observed in electromagnetic radiation, for which such an analysis
is not possible.

\paragraph{Tidal disruption of stars.}
These catastrophes occur when a star collides with a black hole and are
observed primarily as flares in the optical band. Rarely, these
events are accompanied by the formation of a jet. The absence of
spatiotemporal correlations with the population of such sources limits
their contribution to the IceCube astrophysical neutrino flux
to the level of $\lesssim 1\%$ (for processes with the jet formation,
$\lesssim 26\%$) \cite{TDE-1908.08547}. There has been some interest in
the IceCube neutrino event IC191001A with an energy of $\sim
217$~TeV, which coincided in arrival direction with one of the tidal
disruption events, AT2019dsg, although the neutrino came 154 days after
it~\cite{TDE-2005.05340}. The authors of that paper give only an estimate
of the probability of a random coincidence, $\approx 4.8 \times 10^{-3}$,
without a detailed statistical analysis. This coincidence is discussed in
more detail in Ref.~\cite{TDE-2005.06097}. Note that in this case, the
tidal disruption of the star was not accompanied by the formation of a
relativistic jet.

\paragraph{Galactic sources.}
As discussed below in Sec.~\ref{sec:general:pi-mesons}, for sources
in our Galaxy, it is possible to observe high-energy neutrinos along with
associated gamma rays of the same energy range. Gamma-ray astronomy
above 100~TeV is now being intensively developed,
motivated in large part by the neutrino results of IceCube.
The observations of point sources~\cite{Tibet-Crab-100TeV,
HAWC-Crab-100TeV, HAWC-J1825-200TeV, HAWC56-100, HAWC-G106,
LHAASO-12sources} and of diffuse gamma rays from the Galactic plane
\cite{Tibet-GalDiffuse} in this band are worth mentioning. So far, the only
indication of a possible connection of the flare of gamma rays with
energies above 100~TeV from a Galactic source and of neutrinos of the same
energy comes from the Carpet-2 facility (Baksan Neutrino Observatory, INR
RAS), which has recorded~\cite{Carpet-Cocoon} the increased flux,
coinciding with the arrival time of a IceCube neutrino with energy $\sim
154$~TeV, from the region of the Galactic disk in the Cygnus
constellation, which includes the so-called Cygnus Cocoon, a probable
source of the most energetic cosmic photon ever recorded
(1.4~PeV~\cite{LHAASO-12sources}). In Fig.~\ref{fig:CarpetCocoon},
\begin{figure}
\centerline{\includegraphics[width=\columnwidth]{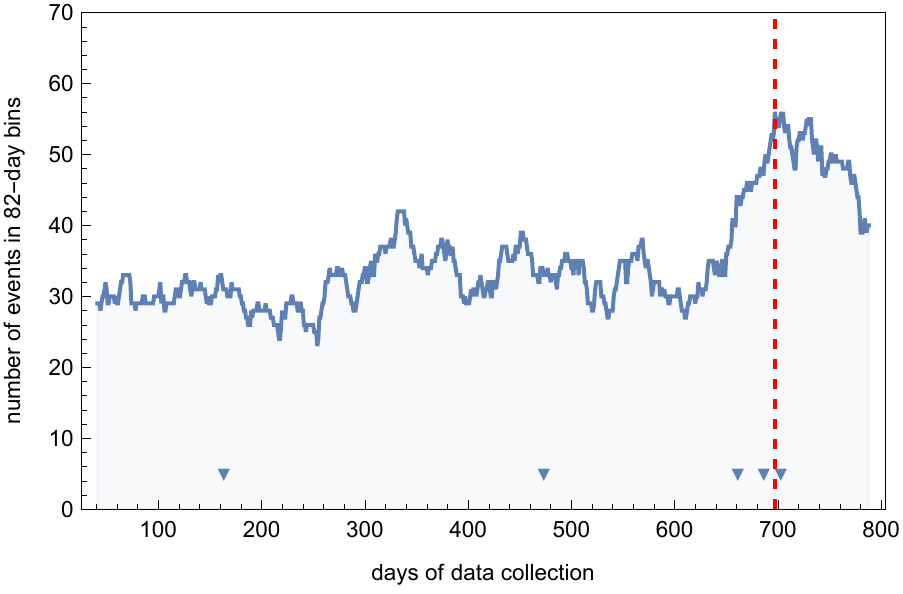}}
\caption{\label{fig:CarpetCocoon}
Count rate of events with estimated primary energy  above 300~TeV
from the direction of Cygnus Cocoon, as recorded by the Carpet-2 facility.
The vertical dashed line indicates the arrival time of a neutrino with
energy $\sim 154$~TeV from this direction. The triangles at the bottom of
the figure indicate the arrival times of the events that have been most
rigorously filtered out as candidates for primary photons. Plotted with
the data from Ref.~\cite{Carpet-Cocoon}.}
\end{figure}
the dependence of the rate of arrival of atmospheric showers from the
direction of this source is shown. The observed excess corresponds to a
flare in photons with energies $\gtrsim 300$~TeV lasting about 3 months
around the day of the neutrino event. The statistical significance of this
flare is $3.1\sigma$ (post trial). Note that although it is the Cygnus
Cocoon that coincides with the the most likely direction of neutrino
arrival, both the accuracy of this direction and the angular resolution of
Carpet-2 are worse than $4^{\circ}$, and other interesting sources are
present in this uncertainty region, including relativistic binary systems
detected in gamma rays above 100~GeV.

\section{General constraints on models of neutrino origin}
\label{sec:general}

\subsection{The $\pi$-meson mechanism and the multimessenger approach}
\label{sec:general:pi-mesons}
Neutrinos have no electric charge and therefore cannot be efficiently
accelerated by external fields. Thus, high-energy neutrinos can only be
produced in decays or interactions of other energetic or heavy particles.
Examples of these processes are decays of ultrarelativistic nuclei (we
will mention this mechanism in the end of this section), decays or
annihilation of very heavy slow particles (dark matter, see
Sec.~\ref{sec:gal}). However, the most natural, guaranteed under
astrophysical conditions mechanism is associated with the decays of
$\pi$-mesons born in the interactions of cosmic rays with matter or
radiation.

Indeed, experiments indicate the presence in our Galaxy (and,
probably in other galaxies as well!) of high-energy hadrons --
cosmic-ray particles, protons and nuclei
\cite{ST-UFN-CR, SemikozKachReview}.
These particles interact with the surrounding matter and radiation. At
high energies, most of the interactions involving hadrons
result in the production of the lightest of strongly interacting
particles, $\pi$ mesons. The latter are unstable and their decay products
include high-energy neutrinos. These
processes must go wherever there are high-energy hadrons (cosmic rays),
but their intensity depends on the amount of matter in the medium or
of the target photons.

\paragraph{Proton-proton interactions.}
Without loss of generality, we will assume that both relativistic hadrons
and ambient hadrons are protons ($p$). At high energies, $pp$-interactions
proceed with the birth of one or more $\pi$ mesons \cite{Berez_pp,
Kelner_pp}, with the leading $\pi$-meson carrying away on average $\sim
1/5$ part of the energy of the relativistic proton, $E_{p}$, in the rest
frame of the target proton. The birth of $\pi^{0}$, $\pi^{+}$ and
$\pi^{-}$ mesons are approximately equally probable, and high-energy
emission is determined by decays (see Fig.~\ref{fig:pi-mesons})
\begin{figure}
\centerline{\includegraphics[width=\columnwidth]{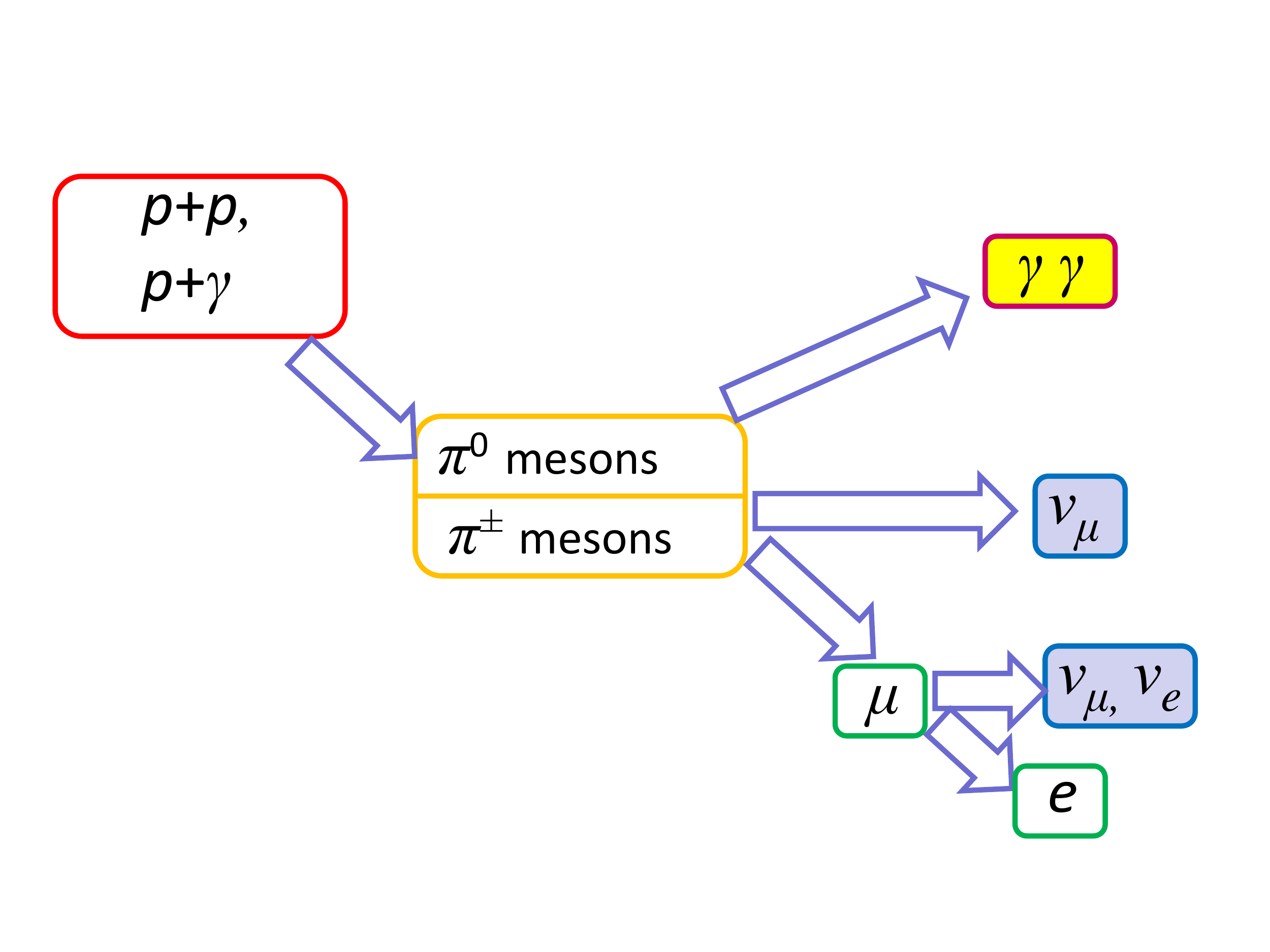}}
\caption{\label{fig:pi-mesons}
Basic processes of neutrino production in astrophysical
sources~\cite{ST-mnogok}.}
\end{figure}
\begin{equation}
\pi^{0} \to \gamma\gamma,
\label{Eq:decay-pi0}
\end{equation}
\begin{equation}
\pi^{+} \to \mu^{+}\nu_{\mu},
\label{Eq:decay-pi+}
\end{equation}
\begin{equation}
\mu^{+} \to e^{+}\nu_{e}\bar\nu_{\mu}
\label{Eq:decay-mu+}
\end{equation}
(the decay chain of $\pi^{-}$ and $\mu^{-}$ is obtained from
Eqs.~(\ref{Eq:decay-pi+}), (\ref{Eq:decay-mu+}) by charge conjugation).
The energy
of a charged $\pi$-meson is divided, on average about equally, among the
four leptons in the final state, or more precisely~\cite{Capone,
Lipari:2007su}
\[
\frac{\langle E_{\nu_{\mu}}\rangle}{E_{\pi^{+}}}
=\frac{1}{2}
\left(1-\left(\frac{m_{\mu}}{m_{\pi}}\right)^{2}   \right)
\simeq 0.21,
\]
\[
\frac{\langle E_{\nu_{e}}\rangle}{E_{\pi^{+}}}
=\frac{1}{10}
\left(2+\left(\frac{m_{\mu}}{m_{\pi}}\right)^{2}   \right)
\simeq 0.26,
\]
\[
\frac{\langle E_{\bar\nu_{\mu}}\rangle}{E_{\pi^{+}}}
=
\frac{\langle E_{e_{+}}\rangle}{E_{\pi^{+}}}
=\frac{1}{20}
\left(3+4\left(\frac{m_{\mu}}{m_{\pi}}\right)^{2}   \right)
\simeq 0.27.
\]
The energy of the neutral $\pi$-meson is divided equally between the two
produced photons. Thus, as a result of the interaction of a proton with
$E_{p}$ with a proton at rest, on average, either two photons
with energies of $E_{p}/10$ each, or (with the probability of 2/3) three
(anti)neutrinos with energies of $E_{p}/20$ are produced.

\paragraph{Proton-photon interactions.}
Consider now the interaction of relativistic protons with photons
of the ambient radiation. At extremely high energies, this process
also proceeds with multiple $\pi$-meson production, and in terms of the
resulting neutrinos is not much different from the $pp$-interaction.
However, we will be more interested in the other regime. Indeed,
significant concentrations of photons in astrophysical sources are
reached only at relatively low energies, so the main contribution to the
production of astrophysical neutrinos comes from interactions with the
single $\pi$-meson production. Unlike for $pp$, the production of
$\pi$ mesons in $p\gamma$ interactions is a threshold process, and the
total energy must be sufficient to produce at least a $\pi$
meson at rest. The main channel for this is the resonant birth and
subsequent decay of the $\Delta^{+}$ baryon,
\begin{equation}
p+\gamma ~\to ~\Delta^{+} ~\to
\left\{
\begin{array}{l}
n + \pi^{+},\\
p+\pi^{0}
\end{array}
\right.
\label{Eq:DeltaResonance}
\end{equation}
($n$ denotes the neutron).

The first thing that draws attention is that in this process, no
leading $\pi^{-}$ is produced, so the equality of $\nu_{e}$ and
$\bar\nu_{e}$ at birth is violated. However, for the sum of neutrinos and
antineutrinos, the flavor content obtained in $p\gamma$-interactions does
not differ from the $pp$ case.

Further, the kinematics of the two-particle $\Delta^{+}$ decay allows us to
determine the fraction of the energy of the initial particles carried away
by the $\pi$-meson: it is again $\sim 1/5$ (the numerical agreement with
the $pp$ case is occasional and takes place only in the $\Delta$-resonance
approximation), so, like for the $pp$-process, $E_{\nu}\sim E_{p}/20$ and
$E_{\gamma}\sim E_{p}/10$ (but, due to the absence of $\pi^{-}$, photons
are now born in about 1/2 of cases rather than 1/3).

Finally, the resonance condition will be written as
$E_{p}E_{\gamma_{B}}=m_{\Delta}^{2}$, where $E_{\gamma_{B}}$ is the energy
of the initial photon, and $m_{\Delta}\approx 1.23$~GeV is the
$\Delta$ baryon mass. Therefore, for the production of a neutrino
with energy $E_{\nu}$ in the $p\gamma$ process, we can estimate the
required energy not only of the proton, $E_{p}\simeq 20 E_{\nu}$, but
also of the initial photon,
\begin{equation}
E_{\gamma_{B}}= \frac{m_{\Delta}^{2}}{20 E_{\nu}} \simeq 750~\mbox{eV}
\left(\frac{E_{\nu}}{100~\mbox{TeV}} \right)^{-1}.
\label{Eq:Egamma-DeltaRes}
\end{equation}
For practical applications, it is important to remember that
Eq.~(\ref{Eq:Egamma-DeltaRes}) is written in the source frame, so
$E_{\nu}$ here may be different from the observed one, e.g., if the source
is located at a cosmological redshift $z$ or if the source is moving
relativistically along the line of sight.
Note also that the estimates given here and in the description of the
$pp$-process are easy to memorize but approximate and in a
number of cases do not work, see, for example, Ref.~\cite{VissaniNot20Ep}.

In many models of acceleration of cosmic-ray particles, the
spectrum of relativistic protons turns out to be power-like. For the
scattering on protons at rest, the scale
invariance of the $pp$-process leads to the fact that the spectra of
secondary particles, and, eventually, neutrinos, repeat the
power-law spectrum of the original protons. The resonant
nature of the $p\gamma$-scattering, in contrast, leads to neutrino spectra
that are different from the power law and are determined by the
spectra of target photons.

\paragraph{Oscillations and flavor content.}
So, from the decays of $\pi^{\pm}$-mesons, one obtains $\nu_{e}$,
$\nu_{\mu}$ and corresponding antineutrinos. In $pp$-interactions,
neutrinos and antineutrinos are born equally, in the $p\gamma$ case the
production of $\bar\nu_{e}$ is suppressed in the $\Delta$-resonance
approximation. In either case, the number of produced $\nu_{\mu}$ and
$\bar\nu_{\mu}$ is twice that of $\nu_{e}$ and $\bar \nu_{e}$.
However, on the way from the source to the
observer, the flavor composition of the neutrinos changes due to
oscillations.

Denote the difference in the squared masses of the two neutrino states as
$\Delta m^{2}$, the neutrino energy as $E_{\nu}$. Then the oscillation
length is
\begin{equation}
L_{\rm osc}=4\pi \frac{E_{\nu}}{\Delta m^{2}} \approx 22.4~\mbox{a.u.}
\left(\!\frac{E_{\nu}}{100~\mbox{TeV}} \!\right)\! \left(\!\frac{\Delta
m^{2}}{7.4\! \times  \! 10^{-5}~\mbox{eV}^{2}} \!\right)^{-1}
\label{Eq:Losc}
\end{equation}
Given that the radius of the orbit of Uranus is about 19~a.u., we obtain
that the distance to any neutrino source outside the Solar system is much
larger than the oscillation length (to estimate this, the smallest
$\Delta m^{2}$ in the system of three neutrino flavors was substituted
into Eq.~(\ref{Eq:Losc})). In addition, for most astrophysical sources
their size is also noticeably larger than $L_{\rm osc}$. Therefore, for
the flavor composition of neutrinos at detection, it makes sense to
consider only the average over the oscillation phases, which means that
the so-called Gribov-Pontecorvo approximation works. Within this
approximation, the fractions of neutrinos of flavors $l=e,\mu,\tau$ at the
observation point, $f_{l}$, are related to those at the source,
$f_{l}^{0}$, by a simple relation,
\begin{equation}
f_l=\sum\limits_{l'}M_{ll'}f_{l'}^{0},
\label{Eq:Gribov}
\end{equation}
where the transformation matrix
$M_{ll'}=\sum_{i}|U_{li}^{2}||U_{l'i}^{2}|$, and $U_{li}$ is the
neutrino mixing matrix (the Pontecorvo-Maki-Nakagawa-Sakata matrix, see
e.g.\ Ref.~\cite{Kudenko-UFN}). The values of elements of this matrix, the
neutrino oscillation parameters, are known with certain
accuracy and continue to be refined in new experiments, see Ref.~\cite{PDG}.
For the purposes of this review, an approximation in which
$f_{\tau}^{0}=0$ is sufficient, then $f_{\mu}^{0}=1-f_{e}^{0}$ and
\begin{equation}
\left\{
\begin{array}{ccl}
f_{e}& \simeq & 0.18+0.36 f_{e}^{0},\\
 f_{\mu}& \simeq & 0.44-0.25
f_{e}^{0},\\
f_{\tau}& \simeq & 0.38-0.11 f_{e}^{0}.
\end{array}
\right.
\label{Eq:flavours}
\end{equation}
Notably, for the value $f_{e}^{0}\approx 1/3$ expected from $\pi$-meson
decays in the source, we obtain an approximate equality of the
detection probabilities of the three neutrino flavors at the detection
point\footnote{The expression (\ref{Eq:Gribov}) is valid for both
neutrinos and antineutrinos. Given that, with the exception of the
Glashow resonance, it is impossible to distinguish $\nu$ and $\bar\nu$ by
modern experiments, we do not distinguish between them in these
expressions.}.

\paragraph{The multimessenger approach.}
Since low-energy photons and target protons are abundant
in many astrophysical objects, the mechanism described above links
fluxes of three types of high-energy particles, the
astrophysical messengers, -- neutrinos from $\pi^{\pm}$ decays, photons
from decays of $\pi^{0}$ and the original relativistic protons, cosmic
rays. This link underlies the so-called multimessenger approach to
observational constraints on source models. Constraints of this type have
been used to estimate high energy astrophysical neutrino fluxes for
decades before their discovery~\cite{BerezinskySmirnov1975, Berez-book,
WaxmanBahcall}, and after the discovery they became a powerful
quantitative tool to understand the neutrino origin.

We have seen that in $\pi$-meson decays, the emitted flux of neutrinos of
energies $E_{\nu}$ is accompanied by an emitted flux of photons at
energies $E_{\gamma}\sim 2E_{\nu}$. The energy fluxes (measured, for
example, in TeV/cm$^2$/s) carried by photons $\gamma$ and neutrinos $\nu$,
are related as
\[
E_{\gamma}^{2} \frac{dN_{\gamma }}{dE_{\gamma }} = A E_{\nu}^{2} \left.
\frac{dN_{\nu }}{dE_{\nu }} \right|_{E_{\nu }=E_{\gamma}/2} ,
\]
where $A\sim 2/3$ ($4/3$) for the $pp$ ($p\gamma$) mechanism, respectively.
However, unlike neutrinos, high-energy photons do not propagate through
the Universe freely: they produce electron-positron pairs
$e^{+}e^{-}$ in interactions with photons of the background
radiation~\cite{Nikishov1962}. This process is a threshold process, and
its cross section peaks strongly just above the threshold, when the
energies of the initial photons satisfy the ratio
$E_{\gamma}E_{\gamma_{B}}\approx 4 m_{e}^{2}$, where $m_{e}$ is the
electron mass. It is convenient to rewrite this relation in
the form
\begin{equation}
\frac{E_{\gamma}}{\mbox{TeV}} \cdot \frac{E_{\gamma_{B}}}{\mbox{eV}}
\approx 1,
\label{Eq:PPenergies}
\end{equation}
from where it follows that photons with $E_{\gamma}\sim (100-1000)$~TeV
produce pairs mostly on cosmic microwave background photons,
whose density in the Universe is very high. As a
consequence, the mean free path of the photons born jointly
with neutrinos does not exceed the size of the Galaxy,
or its immediate vicinity. This, however, is only a part of the story,
since the produced $e^{\pm}$ interact with the same background radiation,
transferring their energy to photons (inverse Compton scattering). The
resulting photons, each with energy already slightly lower than the
energy of the initial one, again produce electron-positron pairs, and
the electromagnetic cascade continues (Fig.~\ref{fig:EMcascade}).
\begin{figure}
\centerline{\includegraphics[width=0.8\columnwidth, trim = 1cm 1cm 10cm
5cm, clip]{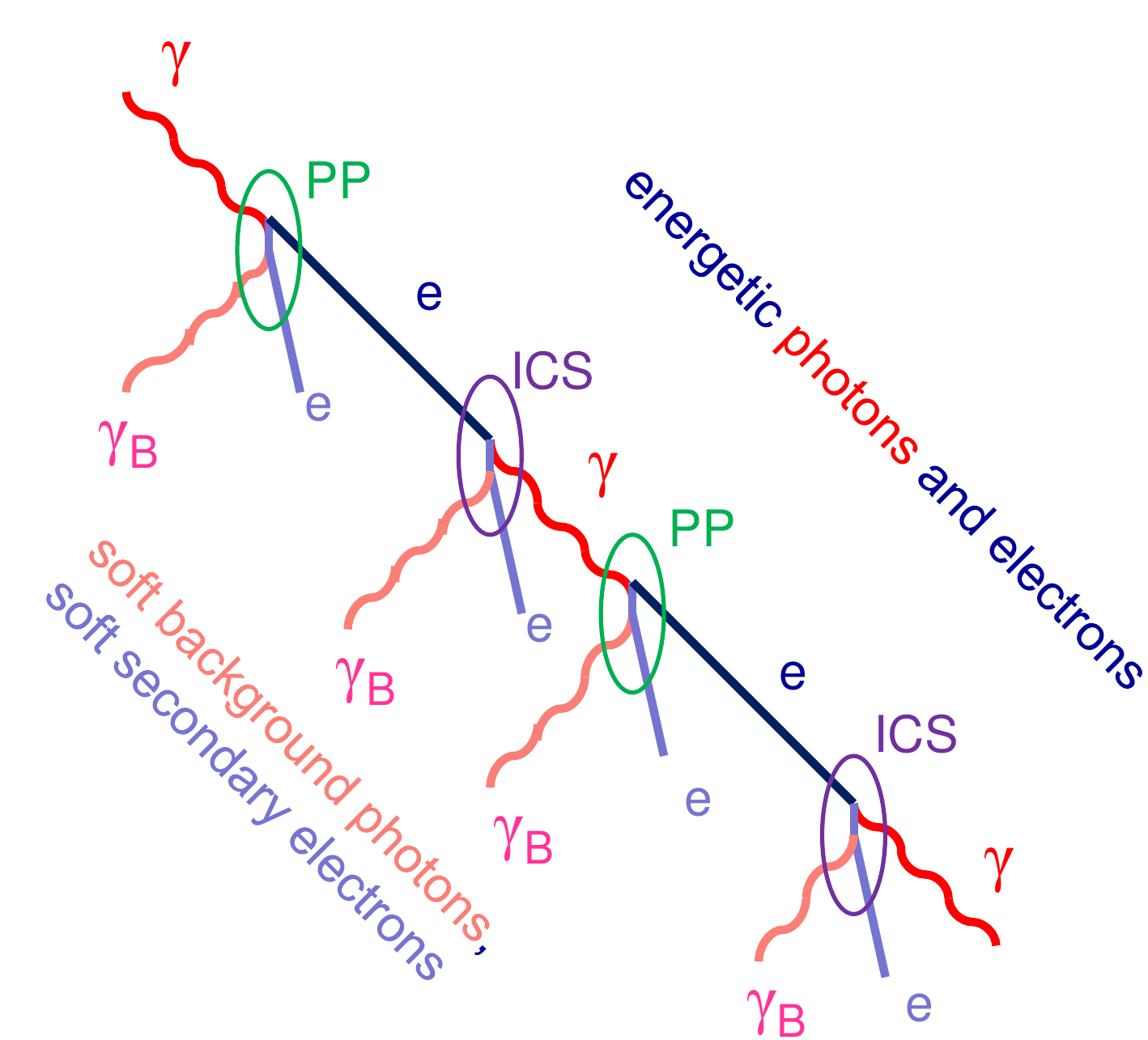}}
\caption{
\label{fig:EMcascade}
The electromagnetic cascade development~\cite{ST-mnogok}. PP -- pair
production, ICS -- inverse Compton scattering. }
\end{figure}
After each cycle, the energy of the original photon is redistributed among
secondary photons, so that the average energy of the photons
in the cascade decreases, although the total energy is conserved (see
Refs.~\cite{Berez-book, BerezinskyKalashev}). This continues until the
average energy of the photons drops so much that they continue moving
through the Universe unimpeded. This occurs at photon energies of several
tens of GeV, so that all energy of the extragalactic diffuse gamma rays
with $E_{\gamma} \gtrsim 100$~TeV is ``pumped'' into the $\sim 10$~GeV
range. The flux of isotropic gamma rays in this range have been
measured \cite{FermiDiffuse} by the LAT instrument aboard the Fermi
satellite, so that the flux obtained by cascading companion photons from
neutrino sources in no way can exceed these measurements. Such a
requirement allows one to set an upper limit on the diffuse neutrino flux
from extragalactic sources~\cite{BerezinskySmirnov1975} unless they
themselves absorb all photons with energies $\gtrsim 10$~GeV. The
quantitative value of this limit depends, although not too much, on
assumptions about the shape of the neutrino spectrum, on the mechanism
($pp$ or $p\gamma$) and, most importantly, on the distribution of sources
in the Universe (evolution). On the other hand, the contribution of
Galactic sources can be estimated from observations of photons with
$E_{\gamma} \sim 2E_{\nu}$, which within the Galaxy can reach the observer
almost without absorption, see Fig.~\ref{fig:Gal-photons}.
\begin{figure}
 \centerline{\includegraphics[width=\columnwidth]{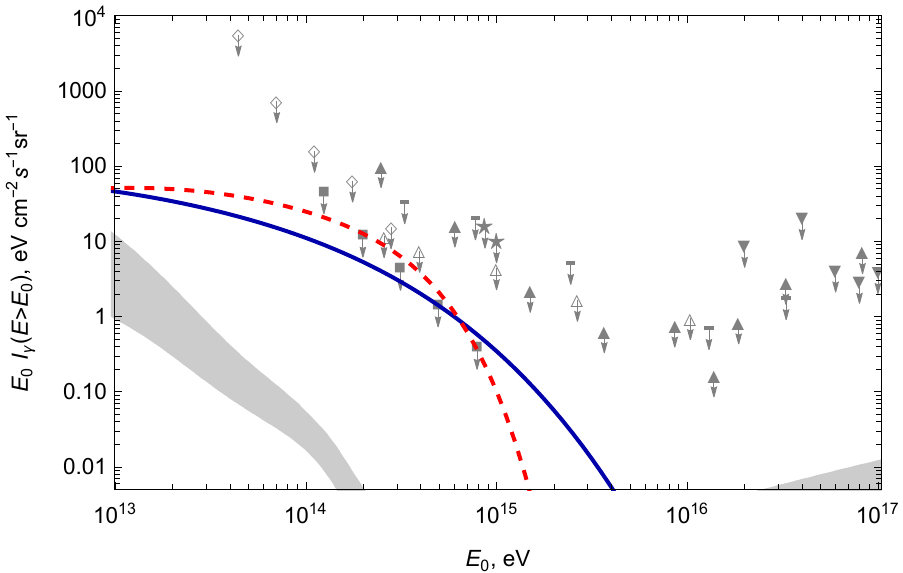}}
\caption{
\label{fig:Gal-photons}
Predictions of the integral isotropic diffuse flux
of high-energy photons in the models explaining the observed neutrino
flux: Galactic (solid line -- interaction of cosmic rays with gas in the
walls of the Local Bubble~\cite{LocalBubble2020}, dashed line --
dark-matter particle decays~\cite{NeronovBubble2018}) and extragalactic
(shaded region~\cite{KalashevST-gamma-Gal}). The gray symbols are
experimental upper limits (empty rhombuses --
GRAPES-3~\cite{GRAPES3-gamma, GRAPES3-spectrum}, empty triangles --
KASCADE~\cite{KASCADE-2003-gamma}, upward triangles -- KASCADE and
KASCADE-Grande~\cite{KASCADE-2017-gamma}, downward triangles --
EAS-MSU~\cite{EAS-MSU-gamma}, horizontal dashes --
CASA-MIA~\cite{CASA-MIA-gamma}, asterisks -- EAS-TOP~\cite{EAS-TOP-gamma},
squares -- preliminary analysis of Tibet-AS$\gamma$
in Ref.~\cite{Neronov:2021ezg}).}
\end{figure}
In the context of IceCube measurements, these possibilities have been
investigated, for example, in the papers~\cite{KalashevST-gamma-Gal,
Murase-gamma-pp-extragal, Murase-gamma-pgamma-extragal, Gupta-Gal,
Winter-gamma-Gal, AhlersMurase-gamma-Gal}.

A similar, though requiring much more assumptions, construction can be
built also for cosmic rays. This widely known estimate
has been called the Waxman-Bahcall limit~\cite{WaxmanBahcall}, although,
as we shall see below, it gives some characteristic value of the flux
rather than constrains it from above.
Suppose that all cosmic rays of ultra-high energy detected
by appropriate detectors are accelerated in some extragalactic
sources, and determine the number of high-energy protons from
measured fluxes at energies $\sim 10^{19}$~eV (at these energies
cosmic rays are guaranteed to be extragalactic in origin). The same
mechanism that accelerates some protons to the highest energies
would accelerate many more protons to lower energies,
$\sim(10^{16}-10^{17})$~eV. In contrast to the most energetic ones, the
protons at lower energies are trapped by the magnetic field in the
source, and sooner or later interact there with matter or radiation,
giving rise to $\pi$ mesons. The neutrino flux that would be produced if
the energy of all these protons were transferred to the $\pi$ mesons,
decaying into photons and neutrinos~\cite{WaxmanBahcall}, gives by
construction an upper limit on neutrino fluxes (this energy may
remain in cosmic rays or be reemitted in other ways). However, soon after
the publication of this estimate, it became clear that it uses too many
assumptions~\cite{Mannheim:not-WB}, first of all, the rather finely tuned
assumption that cosmic rays with energies $\sim 10^{19}$~eV leave the
source, but protons of lower energy, those producing the observed
neutrinos, stay there. It is unclear whether cosmic rays of ultra-high
energies and neutrinos of energies $\sim 100$~TeV must be produced in the
same sources, since to accelerate protons to $\sim 10^{19}$~eV,
much more exotic, difficult to achieve conditions are required than those
necessary for the neutrino-producing protons of $E_{p}\sim 20E_{\nu}\sim
10^{16}$~eV. Finally, there is still no clarity, at which energies cosmic
rays detected at the Earth have the extragalactic origin, while the
numerical value of the obtained estimate depends considerably on this.

The appealing simplicity of these estimates has been one of the
important motivations for building cubic-kilometer
scale neutrino telescopes, whose sensitivity just reaches such flux values
\cite{BerezZatsepin-UFN1977}. Surprisingly, the observed IceCube
neutrino fluxes are, by the order of magnitude, close to these estimates,
understood as upper limits, although originally this
could not have been predicted:
\begin{quote}
``Experiments with extragalactic neutrinos\dots, given a positive result,
will provide very important and unique information both for astrophysics
and for elementary-particle physics. If the result is negative
(with an installation of
$10^{9}$~m$^{3}$), this will merely give cause to the mournful
arguments by the astrophysicists, who will undoubtedly find many
``natural'' explanations for the low flux of extragalactic neutrinos''
(V.S.~Berezinsky, G.T.~Zatsepin~\cite{BerezZatsepin-UFN1977}).
\end{quote}
One might even say that one of the major difficulties in
theoretical explanation of the origin of the detected astrophysical
neutrinos is precisely that their flux turned out to be very large, at
the level of maximal expectations. And these
estimates have remained in the focus also of more detailed studies
of the neutrino origin.

Let us take a look at Fig.~\ref{fig:long-MM}.
\begin{figure*}
\centerline{\includegraphics[width=0.8\textwidth]{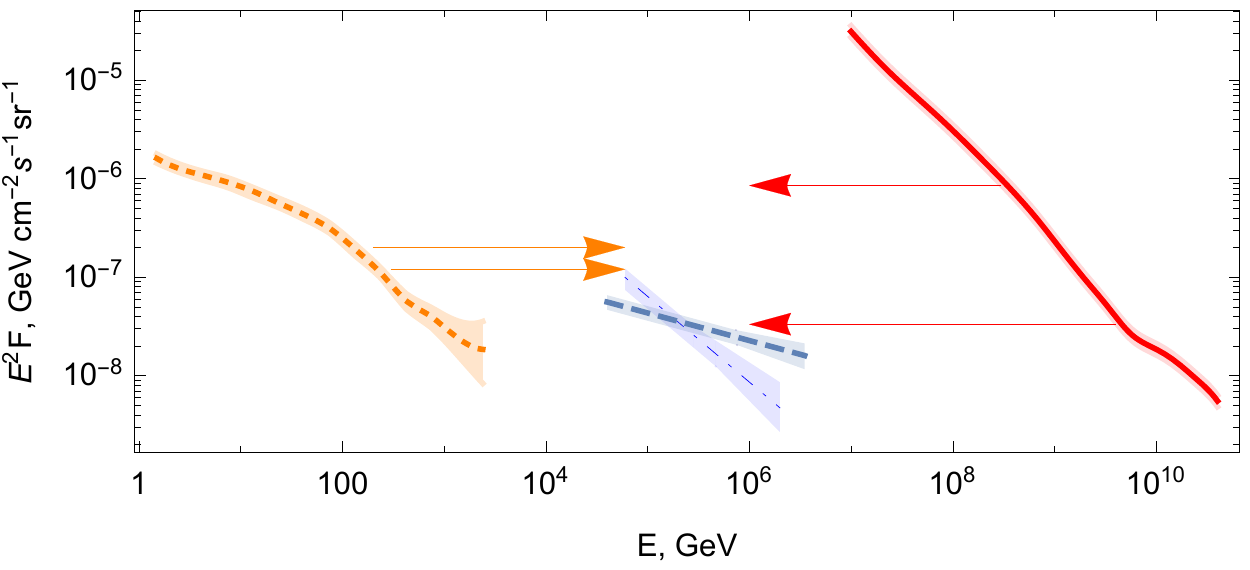}}
\caption{\label{fig:long-MM}
Constraints on and estimates of the neutrino flux in the multimessenger
approach. The orange dashed line on the left, $E\sim (1-1000)$~GeV, --
diffuse isotropic gamma ray flux according to Fermi LAT data
\cite{NeronovSemikozFermiDiffuse}. In the middle, $E\sim
(10^{5}-10^{6})$~GeV, -- total fluxes of astrophysical neutrinos of all
flavors according to IceCube data (blue dashed line -- $\nu_{\mu}$
2019~\cite{IceCube-mu2019}; thin blue dot-dashed line -- HESE
2020~\cite{HESE2020}). On the right, $E\sim (10^{7}-10^{10})$~GeV, --
cosmic-ray flux according to the combined fit~\cite{GSFflux-ICRC2017}. The
arrows illustrate the estimates and constraints discussed in the text and
their uncertainties. }
\end{figure*}
Similar plots are often used to illustrate the
multimessenger approach. It shows the energy fluxes
of diffuse photons, neutrinos, and cosmic rays recorded
by different instruments in the corresponding energy ranges. One can
note that
the fluxes of neutrinos $\sim$~PeV, photons $\sim 10$~GeV, and protons
$\sim 10^{19}$~eV are of the same order of magnitude, $\sim
10^{-8}$~GeV/cm$^{2}$/sec. This allows one to speculate about the
realization of the scenario described above, saturating the
multimessenger constraints (neutrinos are born in extragalactic sources of
ultra-high energy cosmic rays, and the diffuse gamma
rays are the result of electromagnetic cascades triggered by photons born
jointly with the neutrinos). For neutrinos with energies above 100~TeV such
models can indeed \cite{Kachelriess:2017tvs} be proposed.
However, on closer examination, for the neutrino flux of all energies, such
reasoning looks superficial and not necessarily correct:
\begin{enumerate}
 \item
the neutrino fluxes registered by IceCube in the
cascade mode, at energies below $\sim50$~TeV, violate the
Fermi-LAT upper limit and exceed the Waxman-Bahcall
estimate;
\item
a significant fraction of the diffuse photon flux recorded by Fermi
LAT is associated with numerous distant blazars, which are not resolvable
into individual point sources due to imperfections of the instrument
\cite{Fermi-bckgr-blazars}; and these blazars may be neutrino
sources themselves;
\item
the cosmic-ray spectrum decreases rapidly, and the assertions
of approximate equality of energy fluxes are only valid for certain
energies; it is unclear how these energies are singled out compared to the
somewhat smaller, $\sim 3\times 10^{17}$~eV, at which cosmic rays are also
likely extragalactic (for quantitative estimates, see
Ref.~\cite{RouletWB}).
\end{enumerate}
As discussed above, other observational evidence also suggests
that the picture of the neutrino origin is more complex.

\paragraph{Other regimes and mechanisms.}
Let us briefly mention possible variations of the
standard picture based on the $\pi$-meson decays, discussed above, while
remaining within the Standard Model of particle physics.

\textit{The muon-damp regime}. If the decays of $\pi^{\pm}$-mesons
occur in the region of a strong magnetic field, the relatively
long-lived $\mu^{\pm}$ have time to lose a significant fraction of their
energy to synchrotron radiation before decaying. In this case, all of the
high-energy neutrinos are muon neutrinos directly born from $\pi^{\pm}$,
that is, in the formulas (\ref{Eq:flavours}) we should put $f_{e}^{0}=0$.
Oscillations result in a flavor composition different from the standard
one: the average flux of $\nu_{e}$ and $\nu_{\tau}$ is now not equal to
that of $\nu_{\mu}$, but to $\sim 2/3$ of it. The current accuracy
of the determination of composition is still insufficient to reliably
distinguish between these scenarios, see Fig.~\ref{fig:ratio}. This, and
even more exotic, scenarios are discussed, for example, in
Refs.~\cite{WinterFlavours, TamborraFlavours}. The magnetic field, at
which this mode starts to work, can be estimated from the requirement that
the characteristic synchrotron-loss time of the muon is less than its
lifetime, while the muon energy can be related to that of the neutrino. It
turns out that to damp the muons, the field must exceed
\[
B_{\mu\,\rm damp}\approx 60~\mbox{kG}\,
\left(\frac{E_{\nu}}{100~\mbox{TeV}} \right)^{-1}.
\]
Fields of order $10^{4}$~G may be present in the immediate
neighborhoods of supermassive black holes, so further refinement of the
flavor composition of observed neutrinos could isolate or constrain
this class of potential sources \cite{RyabtsevFalvour}. Note that
the number of photons from $\pi^{0}$ decays per high-energy neutrino in
this case is larger than for the usual scenario.

\textit{Note about beta decays}.
The presence of a significant
number of nuclei heavier than protons among cosmic-ray particles, in
principle, allows for the following scenario for the origin of neutrinos.
A nucleus, like a proton, is accelerated to significant energy and then
decays, for example, by photodesintegration. Some of its fragments,
primarily individual neutrons, may be unstable with respect to beta decay.
Neutrinos will be present among the decay products and inherit
part of the kinetic energy of the original nucleus. This part, however, is
small, since the fraction of the neutrino energy relative to the energy of
the neutron is, in the laboratory frame, only $\sim 10^{-3}$. Therefore,
as a rule, the contribution of such a mechanism to the production of
high-energy neutrinos is small compared to that of conventional hadronic
interactions. The threshold for photodesintegration of nuclei is about an
order of magnitude below the threshold for the $\pi$-meson production, so
there is a short energy range in which this process is
significant~\cite{AnchoBetaDecay}. Muon neutrinos are not produced in this
case, $f_{e}^{0}=1$, and the observed composition, according to
Eq.~(\ref{Eq:flavours}), is also depleted in $\nu_{\mu}$, which is in a
poor agreement with the IceCube results, even taking into account their
low precision \cite{HESE2020, BustamanteFlavour2019}.

\subsection{General constraints on source populations}
\label{sec:general:aniso}
Each particular model of the neutrino origin gives its own
predictions of the distribution of arrival directions in the sky and can be
tested either in a combination of specific individual analyses or on the
basis of the results outlined in Sec.~\ref{sec:exper:aniso}. However,
there exist approaches in which, starting with observational data,
it is possible to obtain general constraints on the number and the neutrino
luminosity of the sources from the main population contributing
to the neutrino flux. In the simplest form, mostly used
today, the sources within the population are assumed to be similar; one
must, however, keep in mind that realistic assumptions about natural
diversity, e.g., of the luminosities of sources within the same class, may
significantly alter the results of such an analysis (for a similar
problem for cosmic rays of ultrahigh
energies, see Refs.~\cite{KachelriessSemikozPopulation, PtuskinPopulation,
PtitsynaPopulation}).

\paragraph{Estimate of the number of sources and their neutrino
luminosity.}
It has already been noted above that the absence of
statistically significant individual sources in the neutrino sky indicates
that the observed flux either has a diffuse origin, or is produced by a
large number of astrophysical objects,
each contributing a little to the observed neutrino flux. Here we
give quantitative estimates of these parameters.

Assume that the observed flux is produced by sources with the same
neutrino luminosity and some redshift distribution
(so called cosmological evolution of sources). One can then estimate how
often, for a given experimental exposure, the observation of more than
one (two, three, \dots) neutrinos from the source (``multiplets'') is
expected. Comparing the number of multiplets in the data with that expected
for a random distribution allows one to constrain the density of the number
of sources in the Universe for a fixed evolution. In a situation where the
constraint is due to the lack of observation of significant clusters of
events corresponding to point sources, it is easy to show that the
assumption of the same luminosity leads to the most conservative
constraint. The method of constraining the number of sources based on
the statistics of arrival-direction clustering has been proposed in
Ref.~\cite{DubovskyMultiplets} for ultra-high energy cosmic rays, but
its application to neutrino astronomy is considerably different, see
e.g., Ref.~\cite{LipariRevMultiplets}. The main difference is that cosmic
rays, because of interactions with the background radiation, are collected
from a limited volume of the Universe, and the absence of clusters means a
large number of sources \textit{in that volume}, that is, a \textit{large}
local source density. The neutrinos are collected from all over the
Universe, and the contribution of (in any case numerous) distant sources
dominates. The lack of clusters for neutrinos indicates that there are no
nearby sources standing out against the uniform background of distant and
weak ones, i.e., the large distance to the nearest source or a
\textit{small} local density\footnote{A lack of clusters both in cosmic
rays and in neutrinos therefore suggests that
it would be unlike
to observe
cross-correlations between their arrival directions
\cite{WinterUHECRnuMultiplets}.}. Of course, in the analysis one
have to account for the large number of random clusters from atmospheric
neutrinos. On the other hand, unlike charged cosmic rays, neutrinos
propagate in  straight lines, so the angular size of the cluster is
completely determined by the precision of the reconstruction of arrival
directions.

Knowing the average neutrino luminosity of each source and the
distribution of sources in distances, i.e., the input data for estimating
the statistics of event clusters, it is easy to calculate the
total neutrino flux and to compare it with observations. The combination of
the two requirements, the absence of significant point sources or
clusters of events and explanation of the observed flux, allows one to
constrain the combinations of neutrino luminosity and source number density
for a fixed assumption about their cosmological evolution. Variations of
this approach have been applied to constraining the parameters of neutrino
sources in Refs.~\cite{AhlersMultiplets, MuraseWaxmanMultiplets, HowBright,
IceCube8yrSourcesMultiplets, NeronovSemikozMultiplets,
MuraseMultiplets2019, FinleyMultiplets}. In Fig.~\ref{fig:multi},
\begin{figure}
\centerline{\includegraphics[width=\columnwidth]{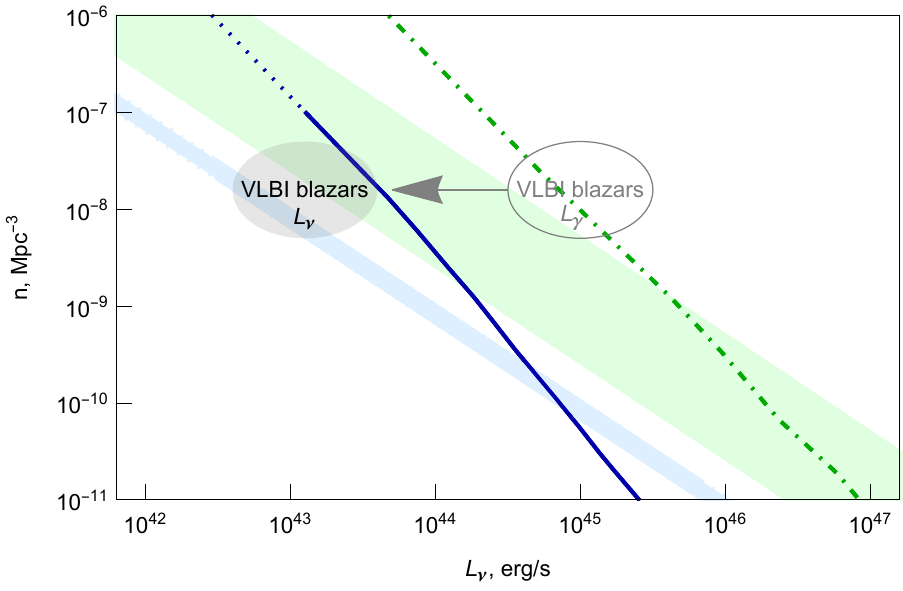}}
\caption{\label{fig:multi}
Constraints on the source number density $n$ (at redshift $z=0$)
and their typical neutrino luminosity $L_{\nu}$. The dashed line --
upper limit (99\% CL) on $n$ from the absence of significant
neutrino point sources in the track sample of 8 years of IceCube data
\cite{FinleyMultiplets}; wide shaded band -- allowed
(99\% CL) region in which these sources explain the flux and spectrum of
IceCube from the same data \cite{FinleyMultiplets}. The solid line
(the dotted line is its extrapolation) is the upper limit (95\% CL) on $n$
from the absence of clusters of neutrino arrival directions in the sample
of IceCube tracks with energies $>200$~TeV
\cite{NeronovSemikozMultiplets}; the narrow shaded band is the allowed
(68\% CL) region in which these sources explain the flux and spectrum
from the same IceCube data \cite{NeronovSemikozMultiplets}. The
constraints assume a strong positive evolution typical for active
galactic nuclei. The shaded ellipse is an estimate of the
characteristic parameters of VLBI-selected blazars, see
Ref.~\ref{sec:exper:aniso:stacking}. The empty ellipse is the same, but
assuming equality of neutrino and bolometric gamma-ray luminosities.}
\end{figure}
results are presented for models with fast positive evolution (many
sources at large redshifts), characteristic for active galactic nuclei.
The constraints for high ($E_{\nu}>200$~TeV) and
all (i.e., mostly $\gtrsim 10$~TeV) energies, obtained from
track events, are shown separately. The combination of the required
neutrino luminosity and the statistics of arrival-direction clusters
places serious constraints on the models of sources. In particular, the
neutrino luminosity of a typical source appears to be substantially smaller
than the characteristic photon luminosity of objects of this type. The
most severe constraints for specific classes of sources arise if, in
addition to clustering, i.e., autocorrelation of neutrino arrival
directions, one takes into account the cross-correlation
of neutrinos and directions to the sources, see
Sec.~\ref{sec:exper:aniso:stacking} \cite{MuraseMultiplets2019}. For
instance, in Fig.~\ref{fig:multi} the region of the characteristic
parameters of VLBI-selected blazars,
Sec.~\ref{sec:exper:aniso:stacking} \cite{neutradio1, neutradio2}, is
shown. To explain the observed flux, their neutrino luminosity must be
of order of a few per cent of the luminosity in photons. It can be seen
that in this case, they can be sources of all astrophysical neutrinos, both
of those with energies above 200~TeV, and with lower energies, in
agreement with the observed directional correlations.

As for the sources with weak evolution,
constraints similar to those shown in
Fig.~\ref{fig:multi} are satisfied only for low
neutrino luminosity and high source number densities. For densities of
$\gtrsim 10^{-6}/$Mpc$^{3}$ these constraints require
to take into account~\cite{Tamborra2MASS} the local large-scale structure of the Universe,
which the distributions of all astrophysical objects follow, when analyzing
neutrino arrival directions; examples of sources of this type are
starburst galaxies. The analyses carried out
\cite{IceCube2MASS, KeFang2MASS} have not yielded strong
constraints on these scenarios yet.

\subsection{Conclusions about general constraints}
\label{sec:general:concl}
Multimessenger analysis of diffuse fluxes of extragalactic cosmic
rays, GeV gamma rays, and high-energy astrophysical
neutrinos demonstrates that the neutrino fluxes detected by
IceCube are at the level of upper limits, i.e., the highest
theoretical expectations, for extragalactic sources with the $\pi$-meson
mechanism. Of course, particular quantitative limits depend on the
details of the model, but qualitatively the picture indicates that the observed
particle fluxes in the three channels -- cosmic rays, neutrinos and
photons, -- are consistent with each other by the order of magnitude. This
allows for the possibility that proton interactions play a significant
role in the origin not only of neutrinos, but also of high-energy gamma
rays. Note that prior to the IceCube observations, the origin of the bulk
of the astrophysical high-energy gamma rays was usually attributed to
relativistic electrons.

At the same time, results of the studies of neutrino arrival directions,
primarily their diffuse nature -- the lack of clearly identified sources
providing for a significant contribution to the flux; the lack of neutrino
correlations with populations of gamma-ray bright potential sources;
strong constraints on the Galactic-disk excess in the neutrino flux, --
point to difficulties of the naive multimessenger approach. It is
not possible to satisfy all constraints simultaneously and to explain all
observations in a simple model with a single class of sources.
It is possible that the observed flux of high-energy astrophysical
neutrinos comes from different classes of sources, which include both
galactic ``PeVatrons'' which dominate at lower energies, and extragalactic
ones (e.g, radio blazars), which provide a harder spectrum. At the same
time, the number of extragalactic
sources contributing to the observed flux is large, and the neutrino
luminosity of each of them is orders of magnitude smaller than the photon
luminosity.

\section{Potential source classes}
\label{sec:sources}
Staying within the frameworks of the general picture derived from
observational constraints and described at the end of the previous section,
we very briefly list here specific astrophysical objects and
environments which are potential sources of observed neutrinos.

\subsection{Models of extragalactic sources}
\label{sec:extragal}

\paragraph{Active galactic nuclei.}
Environments of supermassive black holes in active galactic nuclei are
the most powerful steady sources of radiation in the Universe.
Relativistic blazar jets directed towards the observer result in
the strong enhancement of fluxes of both photons and neutrinos due to
the Lorentzian kinematics.
They have been considered as sources of high-energy neutrinos in
theoretical works since the early days of neutrino
astronomy~\cite{BerezinskyNeutrino77, Eichler1979, BerezGinzb}, see
also recent reviews~\cite{Murase-rev, Meszaros-rev, Boettcher-rev,
Cerruti-rev} and references therein.

The classical scenario~\cite{Sikora1990, Stecker:1991vm, Mannheim1992-FSRQ,
NeronovWhich, Stecker:2013fxa, Dermer-FSRQ, KalashevAGN} involves
the neutrino production in the vicinity of a black hole, closer than
the jet launches. The density of matter in the central regions of galactic
nuclei is low, and the main channel for the neutrino production is
probably the $p\gamma$ process. Relativistic protons with the
required energies up to $\sim (10^{16}-10^{17})$~eV can be accelerated
either in the black-hole magnetosphere \cite{PtitsynaGaps} or in shock
waves in the vicinity of the accretion disk. The latter is also the source
of intense radiation that provides the second necessary ingredient, the
target photons.

In the context of the observational data discussed above, this scenario
requires extension or modification because it does not allow one to
explain the astrophysical neutrinos of energies $\sim 10$~TeV,
for which a connection to blazars has also been
established \cite{neutradio2}. Indeed, according to
Eq.~(\ref{Eq:Egamma-DeltaRes}), production of these neutrinos requires X-ray
target photons, while the accretion disk emits mainly in the ultraviolet
and can provide the target photons for the production of neutrinos of
much higher energies only. Modifications of this scenario include a
contribution from the emission of the accretion disk corona
\cite{InoueCorona1, MuraseCorona1, MuraseCorona2, InoueCorona2} whose
spectrum, compared to the disk itself, extends into higher
energies, although with a lower intensity. A common problem of models of
the neutrino production in the region of the disk and its corona is the
lack of
a direct link to relativistic amplification in
the jets -- as a rule, neutrinos in such models are emitted
isotropically. An additional, albeit avoidable, complication is related
to the fact that, judging by the fast variability, the observed gamma rays
from the blazars originate precisely in this region, while no statistically
significant association between neutrinos and such gamma rays were found.

These problems are circumvented if neutrinos are born already in the
relativistic jets, near their base, see e.g., \cite{neutradio2,
Halzen:1997-jet, AtoyanDermer}. Here, X-ray photons of the target are
present, and the acceleration of protons to the desired energies can occur
in weakly relativistic shock waves \cite{Bykov-acceleration,
Sironi:slow-shocks, LemoineWaxman}. This is where the visible radio
emission of blazars is produced, correlations of neutrino arrival moments
with enhancements in which have been established in \cite{neutradio1,
Hovatta} (regions closer to the black hole are opaque to radio emission).
Since one-zone models fail to describe the entire spectrum of the observed
electromagnetic radiation of blazars from the radio to gamma rays,
it is natural to assume that
the neutrino radiation may not come from the same region as the gamma
radiation. However, it is worth noting that the target photon density
in the jet is small, and to produce the required number of
neutrinos in this region one needs a significant proton power of
the jet \cite{neutradio2}. It could be possible that the observed
correlation between neutrinos and the radio emission indicates the
operation of the $pp$-mechanism \cite{Neronov-radio-pp}.

\paragraph{Cosmic-ray reservoirs.}
This class includes astrophysical objects of various scales
where cosmic rays are trapped for long periods
by  magnetic field. The probability of their interaction with
ambient protons grows with time, so that eventual
neutrino production by the $pp$-mechanism is guaranteed. One
of the most widely discussed source classes of this kind are
starburst galaxies, which have
magnetic fields large enough to confine protons and where mechanisms
to accelerate cosmic protons to high energies may work.
Ref.~\cite{LoebWaxmanStarbursts}, prior to the start of IceCube data
taking, predicted diffuse neutrino flux from a population of
starburst galaxies, at the level of the
subsequently discovered flux. Sources of this class are
prototypical for the Waxman-Bacall estimate, and all the
multimessenger relations and constraints
discussed in Sec.~\ref{sec:general:pi-mesons} work for them. In particular,
such sources are transparent to gamma rays, but are not blazars, so the
``cascade'' limit from Fermi LAT is particularly
serious~\cite{1511.00688noSB, Kistler:noSB}. The reservoirs of
cosmic rays, which in addition to starburst galaxies
include also clusters of galaxies~\cite{Reservoirs-clusters1,
Reservoirs-clusters2} and kiloparsec-scale structures in active
galaxies, radio lobes \cite{Reservoirs-Lobes}, can therefore be
sources of observable neutrinos only with energies $\gtrsim
100$~TeV \cite{Murase:reservoirs}. The interest in this class of sources
is heated \cite{StarburstsStrike} by an excess of IceCube
events from the direction of a nearby powerful starburst galaxy
NGC~1068, see Sec.~\ref{sec:exper:aniso:blindsearch}, but no analysis of
the source population confirms a significant contribution from
cosmic-ray reservoirs to the neutrino flux.

\paragraph{Stopped jets.}
In explosions of very massive stars, not uncommon in particular at the
early stages of the galaxies' evolution, the jets produced in the central
part of the star may not reach the surface because of the high density of
the hydrogen envelope of the star. The result is a ``choked gamma-ray
burst,'' an event with the energy release of a cosmic gamma-ray burst but
without a detectable flare of electromagnetic radiation, which is absorbed
by the outer layers of the star and/or the surrounding matter. In the
process of interaction of the jets with matter, neutrinos are born in
$pp$ interactions~\cite{Meszaros-Chocked2001, Razzaque-Chocked2004,
Ando-Chocked2005, Gupta-Chocked2006}, and because of the lack of visible
gamma-ray bursts, there is no contradiction with the constraints from the
population analysis~\cite{Murase-Chocked2013, Bhattacharya-Chocked2014}.
Theoretical models of such sources contain considerable uncertainties, but
a common feature is a strong peak in the energy distribution of neutrinos
near 100~TeV. As a consequence, such sources cannot explain the entire
observed neutrino flux -- either the observed flux at $\sim 100$~TeV is
explained, but not below or above, or the predicted flux at 100~TeV is too
high \cite{Tamborra-chocked2016, Chocked1512.08513, Chocked1604.08131,
Denton-chocked2018, He-chocked2018, Ahlers-offaxisGRB2019}.

\paragraph{Tidal disruption of stars.}
Falling of a star into a supermassive black hole is a rare event.
On average, in a particular galaxy it happens once per every
$10^{4}-10^{5}$~years. The  star is first
destroyed by tidal forces and then about half of its matter accretes
onto the black hole. A sharp increase in the accretion rate leads in
any case to a flare and, rarely,
to the formation of a jet. One can understand this event as a
short-term transformation of an ordinary galaxy into an active one, so
that neutrinos can be produced in the jet in
the usual $p \gamma$ process. Moreover, the remnants of the star
provide additional target material for the $pp$ process. The
majority of theoretical works on the neutrino production in tidal
disruption events address processes with the
formation of jets~\cite{TDEjet1, TDEjet2, TDEjet3, TDEjet4, TDEjet5,
TDEjet6, TDEjet7}, to which the case of a registered coincidence with the
IceCube event \cite{TDE-2005.05340} does not seem to
belong \cite{TDE-2005.06097, Murase-TDE}. The contribution of tidal
disruption events without jet formation to the observed high-energy
neutrino flux can only be very small, as follows both from a
theoretical point of view \cite{Murase-TDE} and from population
analysis \cite{TDE-1908.08547}.

\subsection{Models of the Galactic flux component}
\label{sec:gal}
When discussing possible scenarios for the origin of neutrinos in our
Galaxy, one should not forget about strict observational constraints
on neutrinos from  the Galactic disk, see Sec.~\ref{sec:exper:aniso:Gal},
Fig.~\ref{fig:KRAgamma}. The constraints on the disk contribution
(and the first indications  of its observations) are at the order of
10\% of the total neutrino flux, in agreement with expectations for the
guaranteed flux from cosmic-ray interactions with the interstellar
gas. Successful scenarios for the origin of the dominant part of the
neutrino flux in the Galaxy must somehow circumvent these disk-related
constraints. There are two possibilities:
\begin{enumerate}
 \item
the flux comes from sources not related to the disk, i.e.
distributed in the Galactic halo (circumgalactic gas or halo
dark matter);
\item
the main contribution to the flux comes from the immediate neighborhood of
the Solar system, that is, from the region smaller in size than the
thickness of the disk.
\end{enumerate}
In addition, the sources can be individual rare objects or
regions in the Galactic plane whose distribution does not follow the
distribution of the disk's gamma rays, which is assumed in the
derivation of constraints of Sec.~\ref{sec:exper:aniso:Gal}.

\paragraph{Circumgalactic gas.}
In recent years, a variety of observational evidence has been obtained
(\cite{Gupta:2012rh}; see e.g. Ref.~\cite{Martynenko:2021aro} and
references therein) for the existence of an extended halo of circumgalactic
gas around our Galaxy, extending up to its virial radius, i.e. about
200~kpc (recall that the radius of the Galactic disk is about 20~kpc,
and its thickness is $<1$~kpc). At energies of $\gtrsim 10^{17}$~eV,
cosmic protons leave the Galactic disk and, interacting with the gas,
can produce high-energy photons \cite{FeldmannCircum}
and neutrinos \cite{AhaOldCircum}. Under the assumption of symmetric
diffusion of cosmic rays and for a realistic model of the circumgalactic
gas, the neutrino flux from such interactions is only a few percent of the
observed one \cite{KalashevCircum}, but it may be possible that
either feeding the halo with cosmic rays through Fermi bubbles, or other
manifestations of past activity in our Galactic nucleus, could alter this
result \cite{AhaNewCircum}.

\paragraph{Decays of dark-matter particles.}
Convincingly confirmed by astronomical observations at various
scales, the existence of invisible matter in the Universe
has yet to be firmly explained in terms of particle physics. There
are many working models, attracting more and more attention while
the scenario of weakly interacting massive particles, most popular for
decades, is gradually being excluded experimentally,
see, e.g.,
the discussion in Ref.~\cite{BertoneTait}. Astrophysical neutrinos can be
produced in the decays of dark-matter particles \cite{DM1, DM2, DM3,
DM4}; to produce decay products of the energies discussed, the dark-matter
particles must be superheavy \cite{SHDM1, SHDM2}. In
the context of explaining the IceCube results, this scenario has been
discussed in particular in Refs.~\cite{DM-ice1, DM-ice2, DM-1703.00451,
DM-1804.04919, DM-1811.04939, DM-1907.11222}. The entire observed
neutrino spectrum from tens of TeV to tens of PeV cannot be explained in
this way, because the energy distribution of the decay products
is noticeably narrower, but this
mechanism can explain the observed flux at some energies. Even for
purely lepton decay channels, very strict constraints on such a scenario
are imposed from the lack of observation of an accompanying flux of
high-energy  photons \cite{DM-1611.08684Misha, DM-1805.04500-Misha,
DM-1905.08662gamma1, DM-2005.04085Misha}.

\paragraph{Gas bubbles and star formation regions.}
In the Galactic disk, neutrinos can be produced in interactions of
cosmic rays with protons and nuclei of the interstellar gas, but
both are  unevenly distributed across the disk. Cosmic rays are
accelerated in sources, ``PeVatrons,'' and are trapped in regions of
high magnetic field, while the distribution of gas is complex because of
stellar winds and shock waves from supernova explosions. As a consequence,
the neutrino signal from the disk may be dominated by the contribution of
a few regions of intense star formation~\cite{Bykov:2015nta,
Bykov-review2015, Bykov:2017fik, Bykov-review2020}, which look like
compact clusters of young massive stars, and so-called superbubbles. A
special place is the Local Bubble, within which our Solar System resides:
the neutrinos coming from it \cite{LocalBubble2020,
NeronovBubble2018, BubbleVelaAniso} do not point back to the Galactic disk
since the size of the bubble, $\sim 100$~pc, is noticeably smaller than
the thickness of the disk. Sources of this type include the Cygnus Cocoon,
mentioned in Sec.~\ref{sec:exper:aniso:flares} in connection with
registration from of a flare of photons with energies above 300~TeV,
coinciding with the IceCube neutrino event \cite{Carpet-Cocoon}. Note that
the available constraints will also be satisfied by neutrinos from other
rare objects, whose distribution in the sky does not follow the profile of
the Galactic gas, such as microquasars in gamma-ray bright binary systems,
one of which coincides, within the accuracy of determining the arrival
direction of neutrinos and photons discussed above, with the
Cygnus  Cocoon  \cite{BykovCygnusPSR}.

\paragraph{Contribution of similar galaxies.}
Speaking of the Galactic origin of neutrinos, we should not forget that
our Galaxy is not unique and sources like these are present not only
in it, but also in billions of other galaxies. Although the contribution
of the sources in our immediate vicinity is significant, neutrinos are
collected from all over the Universe, and the total contribution from
other galaxies may be of the same order as ours. Simple quantitative
estimates \cite{GalloRosso:2018omb} show that this is indeed the fact: by
the order of magnitude, any Galactic contribution to the persistent flux
of high-energy neutrinos would be similar to the total contribution of
similar sources in all other galaxies in the Universe.

\section{Conclusions}
\label{sec:concl}
\begin{itemize}
 \item
Astrophysical neutrinos of high, 10~TeV -- 10~PeV, energies have been reliably
detected by the IceCube experiment; their observation is being confirmed by
the  ANTARES results and by the first Baikal-GVD data. The distribution
in zenith angles of events passing the most strict selection
excludes their atmospheric origin even for exotic assumptions. In
the energy range above 100~TeV, neutrino astronomy has surpassed photon
astronomy and motivated its development.
\item
Although the sources of high-energy astrophysical neutrinos
have not yet been definitively determined, their total fluxes point to a
significant role of relativistic hadrons in high-energy astrophysics:
within standard physics, only processes involving them can give rise to
neutrinos of this energy range.
\item
Spectra of astrophysical neutrinos reconstructed from
IceCube cascade and track events are poorly consistent with each other
under the assumption of a power-law dependence of the flux on the energy.
This can be explained by a more complex shape of the spectrum, reflecting
the combined contribution of different populations of sources. It is
likely that the sky in neutrinos looks no less complex and diverse as in
photons.
\item
The flux of astrophysical neutrinos at energies above $\sim 100$~TeV is
probably dominated by the contribution of numerous distant
extragalactic sources. Population studies indicate a statistically
significant association of neutrinos with blazars, that is the active
galactic nuclei with relativistic jets directed toward the
observer, manifesting themselves by powerful radiation from parsec scales
visible with VLBI. These sources are not always bright in the gamma-ray
band, and their neutrino luminosity on average is noticeably lower than
the photon luminosity. They contribute significantly to the neutrino flux
also at lower energies.
\item
The neutrino flux component that dominates at energies
$\sim(10-100)$~TeV may either be of a Galactic origin or
be connected to multiple extragalactic sources which are opaque to
photons at $\gtrsim$GeV energies. There are observational indications to
the presence of a Galactic component.
\item
The prospects for further
understanding of the nature of high-energy astrophysical neutrinos are
related both to the work of cubic-kilometer scale neutrino telescopes
(including Northern-hemisphere water detectors, Baikal-GVD, which has just
started to work, and KM3NeT which is under construction) and to the
multimessenger analysis, including observations in the electromagnetic
channel in all bands -- from radio (in which, due to the best angular
resolution, it is possible to study blazar jets in the immediate vicinity
of supermassive black holes) to PeV (this emerging field of astronomy will
answer questions about Galactic sources of neutrinos).
\end{itemize}

The author is indebted for interesting and helpful discussions of various
aspects related to the origin of high-energy astrophysical neutrinos to
his colleagues,
M.~Barkov,
A.~Bykov,
H.~Dembinski,
T.~Dzhatdoev,
Zh.-A.~Dzhilkibaev,
G.~Domogatsky,
A.~Franckowiak,
O.~Kalashev,
Yu.A.\ and
Yu.Yu.~Kovalev,
M.~Kuznetsov,
T.~Montaruli,
K.~Murase,
A.~Neronov,
A.~Plavin, E.~Podlesny, K.~Postnov, V.~Rubakov, G.~Rubtsov, K.~Ryabtsev,
D.~Samtleben, D.~Semikoz,
K.~Spiering,
O.~Suvorova
and
K.~Zhuravleva.

This work is supported by the Ministry of Science and
Higher Education of the Russian Federation,
Contract 075-15-2020-778 of the Program of Major Scientific Projects
within the National Project ``Science''.

\bibliography{try-e}
\bibliographystyle{apsrev4-2}
\end{document}